\documentclass[journal]{IEEEtran}

\usepackage{graphicx}
\usepackage{xcolor}
\usepackage{colortbl,booktabs}
\usepackage{amssymb}
\usepackage{amsfonts}
\usepackage{amsmath}
\usepackage{epsfig}
\usepackage{color}
\usepackage{fancybox}
\usepackage{textcomp}
\usepackage{multirow}
\usepackage{setspa ce}
\usepackage{psfrag}
\usepackage{booktabs}
\usepackage{float}
\usepackage{setspace}
\def\bm#1{\boldsymbol{#1}}

\usepackage{color}

\begin{document}
	%
	\title{Deep Learning for Wireless Physical Layer: Opportunities and Challenges}
	%
	%
	%
	
	\author{Tianqi~Wang*,~Chao-Kai~Wen,~Hanqing~Wang,~Feifei~Gao,~Tao~Jiang~and~Shi~Jin
	  
	  * The corresponding author, email: wangtianqi@seu.edu.cn	
		\thanks{T. Wang, H. Wang, and S. Jin are with the National
			Mobile Communications Research Laboratory, Southeast University, Nanjing
			210096, China (e-mail: wangtianqi@seu.edu.cn; hqwanglyt@seu.edu.cn; jinshi@seu.edu.cn).}
		\thanks{C.-K. Wen is with the Institute of Communications Engineering, National
			Sun Yat-sen University, Kaohsiung 80424, Taiwan (e-mail: ckwen@ieee.org).}
		\thanks{F. Gao is with the State Key Laboratory of Intelligent Technology
			and Systems, Tsinghua National Laboratory for Information Science and Technology, Department of Automation, Tsinghua University, Beijing 100084, China (e-mail: feifeigao@ieee.org).}
		\thanks{T. Jiang is with the School of Electronic Information and Communications, Huazhong University of Science and Technology, Wuhan 430074, P. R. China. (e-mail: taojiang@hust.edu.cn).}}	
	\maketitle
	
	\begin{abstract}
		Machine learning (ML) has been widely applied to the upper layers of wireless communication systems for various purposes, such as deployment of cognitive radio and communication network. However, its application to the physical layer is hampered by sophisticated channel environments and limited learning ability of conventional ML algorithms. Deep learning (DL) has been recently applied for many fields, such as computer vision and natural language processing, given its expressive capacity and convenient optimization capability. The potential application of DL to the physical layer has also been increasingly recognized because of the new features for future communications, such as complex scenarios with unknown channel models, high speed and accurate processing requirements; these features challenge conventional communication theories. This paper presents a comprehensive overview of the emerging studies on DL-based physical layer processing, including leveraging DL to redesign a module of the conventional communication system (for modulation recognition, channel decoding, and detection) and replace the communication system with a radically new architecture based on an autoencoder. These DL-based methods show promising performance improvements but have certain limitations, such as lack of solid analytical tools and use of architectures that are specifically designed for communication and implementation research, thereby motivating future research in this field.
		
	\end{abstract}
	

	%
	\IEEEpeerreviewmaketitle

	\section{Introduction}
	\IEEEPARstart{W}{ireless} communication technologies have experienced an extensive development to satisfy the applications and services in the wireless network. The explosion of advanced wireless applications, such as diverse intelligent terminal access, virtual reality, augmented reality, and Internet of things, has propelled the development of wireless communication into the fifth generation to achieve thousandfold capacity, millisecond latency, and massive connectivity, thereby making system design an extraordinarily challenging task. Several promising technologies, such as massive multi-input multi-output (MIMO), millimeter wave (mmWave), and ultra-densification network (UDN) have been proposed to satisfy the abovementioned demands. These technologies demonstrate the same characteristic (i.e., the capability to handle large wireless data). However, extant conventional communication theories exhibit several inherent limitations in fulfilling the large data and ultra-high-rate communication requirements in complex scenarios, listed as follows.

	\subsubsection{Difficult channel modeling in complex scenarios}
	The design of the communication systems significantly depends on practical channel conditions or is based on channel models that characterize real environments implicitly for mathematical convenience. These models struggle in complex scenarios with many imperfections and nonlinearities \cite{O2017An}, although they may capture some features in conventional channels. For example, the increased number of antennas in massive MIMO systems has changed channel properties \cite{larsson2014massive}, and the corresponding channel models remain unknown. The use of out-of-band signals or sensors as sources of side information at mmWave is promising \cite{Prelcic17ArXiv}. However, a method for combining out-of-band and sensor information to obtain the channel state information of the mmWave remains unknown. In scenarios, such as underwater or molecular communications \cite{farsad2017detection}, the channels cannot be characterized by using rigid mathematical models.
	Thus, systems or algorithms that can complete communication tasks without defined channel models are essential.
	
	\subsubsection{Demand for effective and fast signal processing}
	The use of low-cost hardware, such as low-resolution analog-to-digital converters with low energy consumption \cite{jeon2016supervised, tanweiqiang}, introduces additional nonlinear imperfections that require using highly robust receiving processing algorithms (e.g., algorithms for channel estimation and detection). However, using these algorithms may increase computational complexity. Traditional algorithms, such as algorithms for MIMO data detection, are iterative reconstruction approaches \cite{samuel2017deep} that form a computational bottleneck in real time, whereas the real-time large data processing capabilities are necessary for the advanced systems (e.g., massive MIMO, mmWave, and UDN). Therefore, the corresponding algorithms require parallel signal processing architecture \cite{larsson2014massive} to achieve efficiency and accuracy.
	
	\subsubsection{Limited block-structure communication systems}
	Conventional communication systems that are constructed in a divide-and-conquer manner, consist of a series of artificially defined signal processing blocks, such as coding, modulation, and detection; these systems solve the communication problems in imperfect channels by optimizing each block independently. An optimal performance in the entire communication task cannot be guaranteed, although researchers have attempted to optimize the algorithms of each processing module for many years and achieved success in practice, because the fundamental problem of communication depends on the reliable message recovery at the receiver side after the message is sent by a transmitter and traverses a channel \cite{dorner2017deep}. This process does not require a block structure. Therefore, it holds promise for further improvement if the suboptimization for each module is replaced by optimizing for end-to-end performance.
	
	Machine learning (ML) has recently regained attention because of the successful applications of deep learning (DL) in computer vision (CV), automatic speech recognition (ASR), and natural language processing (NLP). The book \cite{Goodfellow-et-al-2016} covers the recent results. Researchers are actively attempting to extend these technologies to other domains, including wireless communication. Embedding ML theories on a wide range of communication systems has had an extensive history and has achieved several successes, especially in the upper layers, such as in cognitive radio, resource management \cite{challita2017proactive}, link adaptation \cite{daniels2010adaptation,pulliyakode2017reinforcement}, and positioning \cite{vieira2017deep}. In contrast to the abovementioned straightforward applications, ML faces several challenges when applied to the physical layer.
	
	Researchers have applied ML to the physical layer for modulation recognition \cite{fehske2005new,azzouz1996modulation}, channel modeling and identification \cite{ibukahla1997neural,sjoberg1995nonlinear}, encoding and decoding \cite{bruck1989neural,ortuno1992error}, channel estimation \cite{wen2015channel}, and equalization \cite{chen1990adaptive,cid1996digital} (see further details in \cite{ibnkahla2000applications} and \cite{jiang2017machine}); however, ML has been unused commercially because handling physical channels is a complex process, and conventional ML algorithms have limited learning capacity. Researchers believe that ML can achieve further performance improvements by introducing DL to the physical layer.  DL possesses essential characteristics, such as deep modularization, which significantly enhances feature extraction and structure flexibility, compared with conventional ML algorithms. In particular, DL-based systems can be used instead of manual feature extraction to learn features from raw data automatically and adjust the model structures flexibly via parameter tuning to optimize end-to-end performance. The DL-based communication system has promising applications in complex scenarios for several reasons.
	
	First, the deep network has been proven to be a universal function approximator \cite{hornik1989multilayer} with superior algorithmic learning ability despite the complex channel conditions \cite{O2017An}. The ``learned'' algorithms in DL-based communication systems are represented by learned weights in DL models that optimize end-to-end performance through convenient training methods instead of requiring well-defined mathematic models or expert algorithms that are solidly based on information theories.
	
	Second, handling large data is an essential feature of DL because of its instinctive nature of distributed and parallel computing architectures, which ensure computation speed and processing capacity. The DL system demonstrates a remarkable potential in producing impressive computational throughput through fast-developing parallelized processing architectures such as graphical processing units.
	
	Third, DL-based communication systems can break the artificial block structure to achieve global performance improvement because they are trained to optimize end-to-end performance without making an implicit request for block-structure. Besides, various libraries or frameworks, such as TensorFlow, Theano, Caffe, and MXNet, have been established to accelerate experiments and deploy DL architectures given the wide application of DL technologies.
	
	Recent studies on DL for wireless communication systems have proposed alternative approaches to enhance certain parts of the conventional communication system (e.g., modulation recognition \cite{O2017An}, channel encoding and decoding \cite{nachmani2016learning,nachmani2017rnn,gruber2017deep,cammerer2017scaling,nachmani2017deep,liang2017iterative}, and channel estimation and detection \cite{samuel2017deep,farsad2017detection,ye2017power,neumann2017learning}) and to replace the total system with a novel architecture on the basis of an autoencoder \cite{O2017An,o2017deep}. This paper aims to provide an overview of these recent studies that focus on the physical layer. We also aim to highlight the potentials and challenges of the DL-based communication systems and offer a guideline for future investigations by describing the motivation, proposed methods, performance, and limitations of these studies. Table \ref{abbreviation} provides the abbreviations appeared in the paper.
	\begin{table}[H]
		\caption{List of abbreviation}
		\label{abbreviation}
		\centering
		\begin{tabular}{c|c}
			\hline\hline
			\bf Abbreviation & \bf Stands for\\
			\hline
			ML & machine learning\\
			\hline
			DL & deep learning\\
			\hline
			MIMO & multi-input multi-output\\
			\hline
			mmWave & millimeter wave\\
			\hline
			UDN & ultra-densification network\\
			\hline
			CV & computer vision\\
			\hline
			ASR & automatic speech recognition\\
			\hline
			NLP & natural language processing\\
			\hline
			SVM & support vector machine\\
			\hline
			NN & neural network\\
			\hline
			GD & gradient descent\\
			\hline
			DNN & deep neural network\\
			\hline
			ReLU & rectified linear units\\
			\hline
			SGD & stochastic gradient descent\\
			\hline
			CNN & convolutional neural network\\
			\hline
			RNN & recurrent neural network\\
			\hline
			LSTM & long short-term memory\\
			\hline
			SNR & signal-to-noise ratio\\
			\hline
			LLR & log-likelihood ratio\\
			\hline
			BP & belief propagation\\
			\hline
			HDPC & high-density parity check\\
			\hline
			BCH & Bose-Chaudhuri-Hocquenghem\\
			\hline
			AWGN & additive whit Gaussian noise\\
			\hline
			BER & bit-error rate\\
			\hline
			mRRD & modified random redundant iterative decoding\\
			\hline
			NND & neural network decoder\\
			\hline
			MAP & maximum a posteriori\\
			\hline
			PNN & partitioned neural network\\
			\hline
			FC & fixed channel\\
			\hline
			VC & varying channel\\
			\hline
			AMP & approximate message passing\\
			\hline
			SDR & semidefinite relaxation\\
			\hline
			ISI & inter symbol interference\\
			\hline
			CP & cyclic prefix\\
			\hline
			MMSE & minimum mean square error\\
			\hline
			RTN & radio transformer network\\
			\hline
			BLER & block error rate\\
			\hline
			CSI & channel state information\\
			\hline\hline
		\end{tabular}
	\end{table}
	The rest of this paper is organized as follows. Section \ref{DL basic} provides a brief overview of the basic concepts of DL. Section \ref{application} presents several application examples of using DL as alternatives for communication systems, such as modulation recognition, channel decoding, and detection. Section \ref{application autoencoder} introduces a novel communication architecture based on an autoencoder. Section \ref{future} discusses the areas for future research. Section \ref{conclusion} concludes the paper.

	\begin{figure}
		\centering
		\includegraphics[width=2.0in]{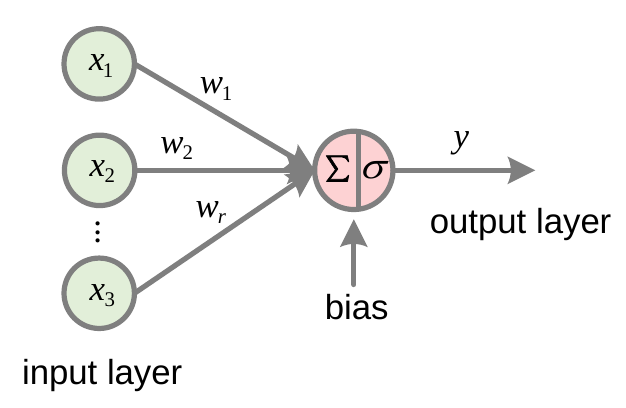}
		\centering\caption{A mathematical model of a neuron, where $r$ features are connected by weighted edges and fed into a sigmoid activation function to generate an output $y$.}
		\label{neural}
	\end{figure}
	
	\begin{figure}
		\centering
		\includegraphics[width=3.0in, height=1.6in]{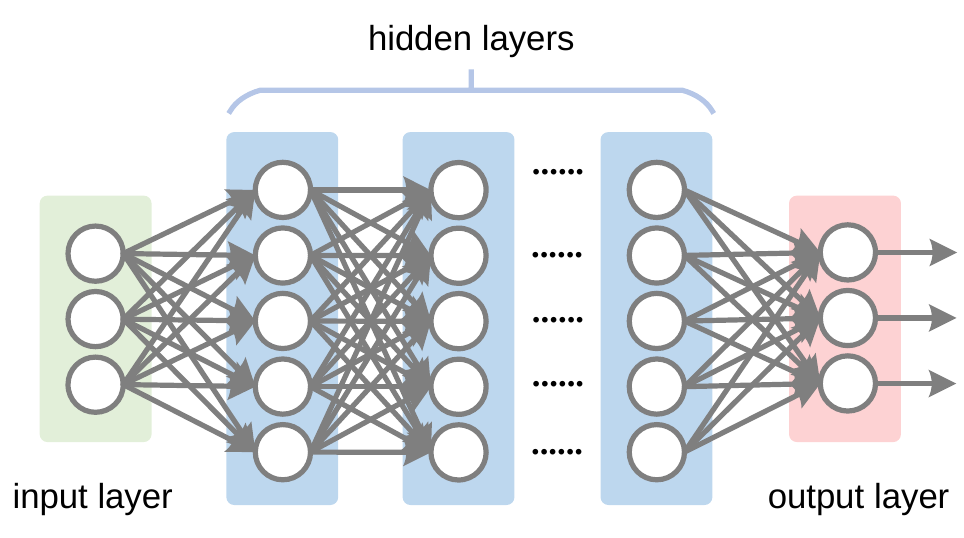}
		\centering\caption{A fully connected feedforward NN architecture where all neurons between adjacent layers are fully connected. The DNN uses multiple hidden layers between the input and output layers to extract meaningful features.}
		\label{DNN}
	\end{figure}
	
	\section{Basic Idea of DL}
	\label{DL basic}
	Langley (1996) defined ML as a branch of artificial intelligence that aimed to improve performance by experience. After long-standing research from the 20th century, various algorithms have been proposed, such as logistic regression, decision tree, support vector machine (SVM), and neural network (NN). As an emerging outstanding algorithm belonging to NN, DL is originally derived from the neuron model that simulates biological neural system schemes as shown in Fig. \ref{neural}. The weighted sum of several inputs with bias is fed into an activation function $\sigma(\cdot)$, usually a sigmoid function, to obtain an output $y$. An NN is then established by connecting several neuron elements to a layered architecture.
	
	The simplest NN is called perceptron, which comprises one input layer and one output layer. A certain loss function, such as square error or cross entropy, must be established for the perceptron to produce a value that is close to the expected one as much as possible. Gradient descent (GD) is commonly used in training the best parameters (i.e., weights and biases) to minimize such loss function. Despite only solving linearly separable problems, a single perceptron will introduce nonlinear properties, and function as a universal function approximator by adding hidden layers and neurons between the input and output layers. The evolved architecture is called multi-layer perceptron that does not significantly differ from the current deep NN (DNN) or NNs with multiple hidden layers, which have achieved success in CV, ASR, and NLP.
	
	A basic DL model is a fully connected feedforward NN (Fig. \ref{DNN}), where each neuron is connected to adjacent layers and no connection exists in the same layer. The back propagation algorithm has been proposed as an efficient method for training such network with GD for optimization. However, the increase in the number of hidden layers and neurons implies the existence of several other parameters to be determined, thereby making the network implementation difficult. Numerous problems may be encountered during the training process, such as vanishing gradients, slow convergence, and falling into the local minimum.
	
	\begin{table}
		\caption{Activation functions}
		\label{activation}
		\centering
		\begin{tabular}{c|c}
			\hline			
			\hline
			\bf Name & \bf Activation function $\sigma(x)$\\
			\hline
			sigmoid & $\dfrac{1}{1+e^{-x}}$\\
			\hline			
			tanh & $\tanh(x)$\\
			softmax & $ \dfrac{e^{x_{i}}}{\sum_{j}e^{x_{j}}} $\\
			\hline			
			ReLU & $\max(0,x)$\\
			\hline\hline
		\end{tabular}
	\end{table}
	
	To solve the vanishing gradient problem, new activation functions (e.g., rectified linear units (ReLU), a special feature of Maxout) have been introduced to replace the classic sigmoid function. Table \ref{activation} shows the activation functions that can adapt to other situations. To achieve faster convergence and decrease computation complexity, classic GD is adjusted into stochastic GD (SGD), which randomly selects one sample to compute the loss and gradient every time. Stochasticity causes severe fluctuation in the training process; thus, mini-batch SGD (a batch of samples computed simultaneously) is adopted as a tradeoff between classic GD and SGD. However, such algorithms still converge to the local optimal solution. To solve this problem and further increase the training speed, several adaptive learning rate algorithms (e.g., Adagrad, RMSProp, Momentum, and Adam) have been proposed. Although the trained network performs well in the training data, this network may perform poorly in the testing process because of overfitting. In this case, early stopping, regularization, and dropout schemes have been proposed to achieve favorable results in training and testing data.
	
\begin{figure}
		\centering
		\includegraphics[width=3.5in]{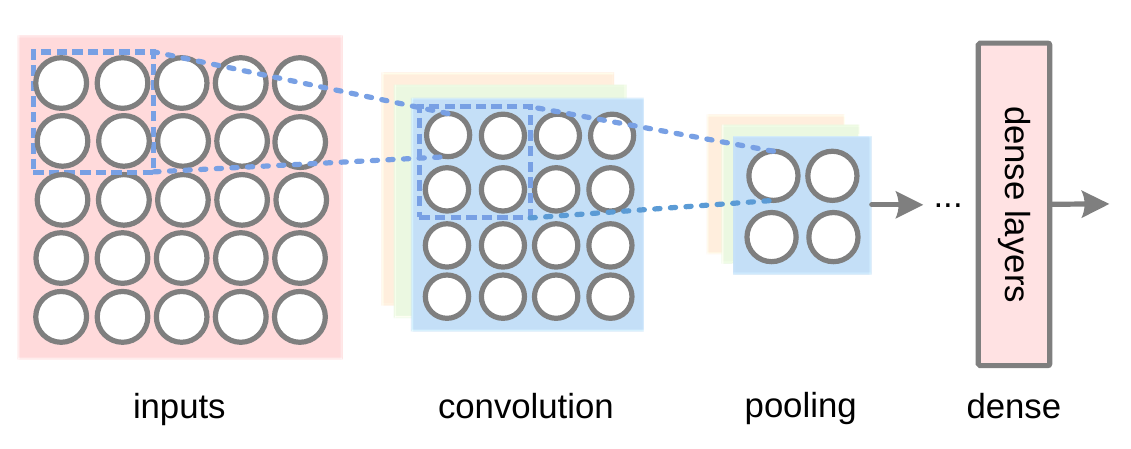}
		\centering\caption{A CNN architecture that adds convolution layers and pooling layers before dense layers. Each output in the convolution layers is obtained by dot production between a certain filter matrix and an input matrix comprised of several neurons in the upper layer. Each output in the pooling layers is obtained by averaging or searching maximum in a group of neurons in the convolution layer.}
		\label{CNN}
	\end{figure}

	Convolutional NN (CNN) is another emerging DNN architecture that is developed from a fully connected feedforward network to prevent a rapid growth in the parameters when applying the latter to image recognition. CNN introduces the idea of designing particular DNN architectures depending on the requirements of specific scenarios. The basic concept of CNN is to add convolutional and pooling layers before feeding into a fully connected network (Fig. \ref{CNN}). In the convolutional layer, each neuron only connects to parts of neurons in the former adjacent layer. These neurons that are organized in a matrix form comprise several feature maps, and the neurons share the same weights in each map. In the pooling layer, the neurons in the feature maps are grouped to compute for the mean value (average pooling) or maximum value (max pooling). Thus, the parameters are substantially decreased before using the fully connected network.
	
\begin{figure}
		\centering
		\includegraphics[width=3.5in]{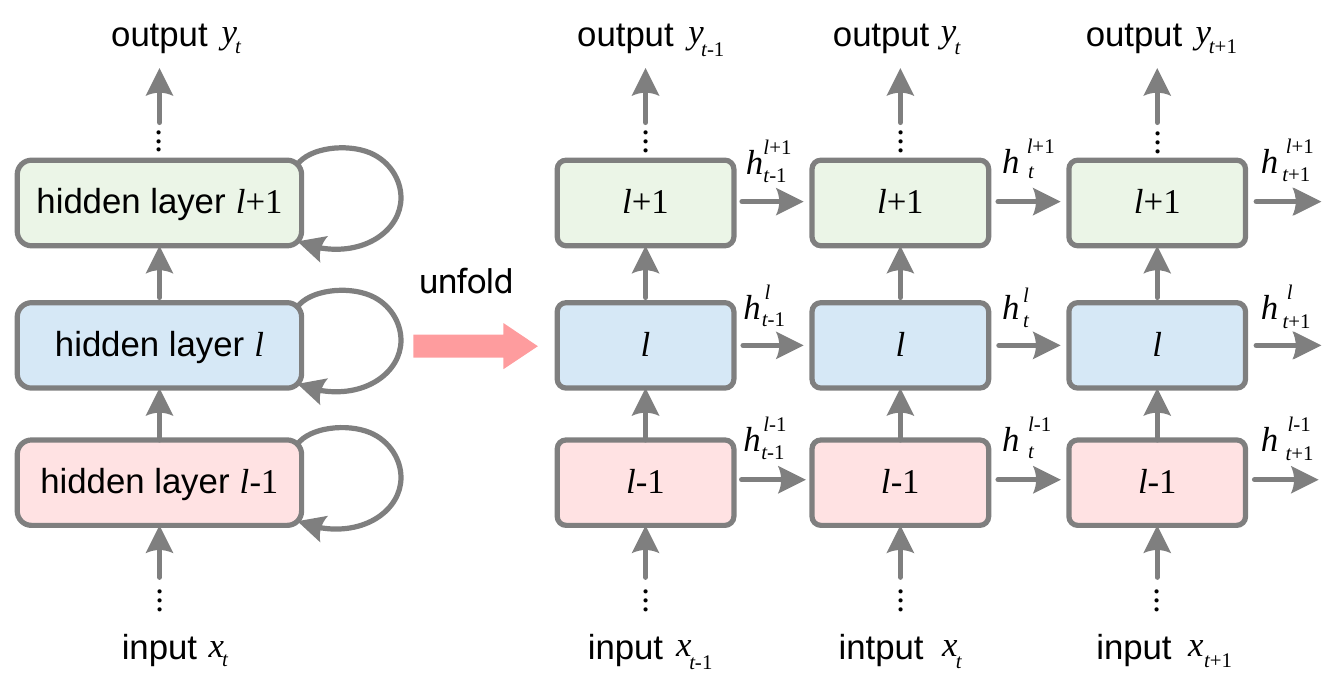}
		\centering\caption{A RNN architecture that considers the extracted features in the former state as the one of the current input information. The current outputs depend on current and former inputs so that the network achieves memory. }	
		\label{RNN}
	\end{figure}

	Recurrent NN (RNN) aims to provide NNs with memory because the outputs depend not only on the current inputs but also on the formerly available information in cases such as NLP. Compared with the aforementioned memoryless NNs without connections in the same hidden layer, the neurons are connected such that the hidden layers consider their former outputs as current inputs to acquire memory (Fig. \ref{RNN}). Some commonly used RNNs include Elman network, Jordan network, bidirectional RNN, and long short-term memory (LSTM).
	
	\section{DL as an Alternative}
	\label{application}
	
	\begin{figure*}
		\centering
		\includegraphics[width=5.5in]{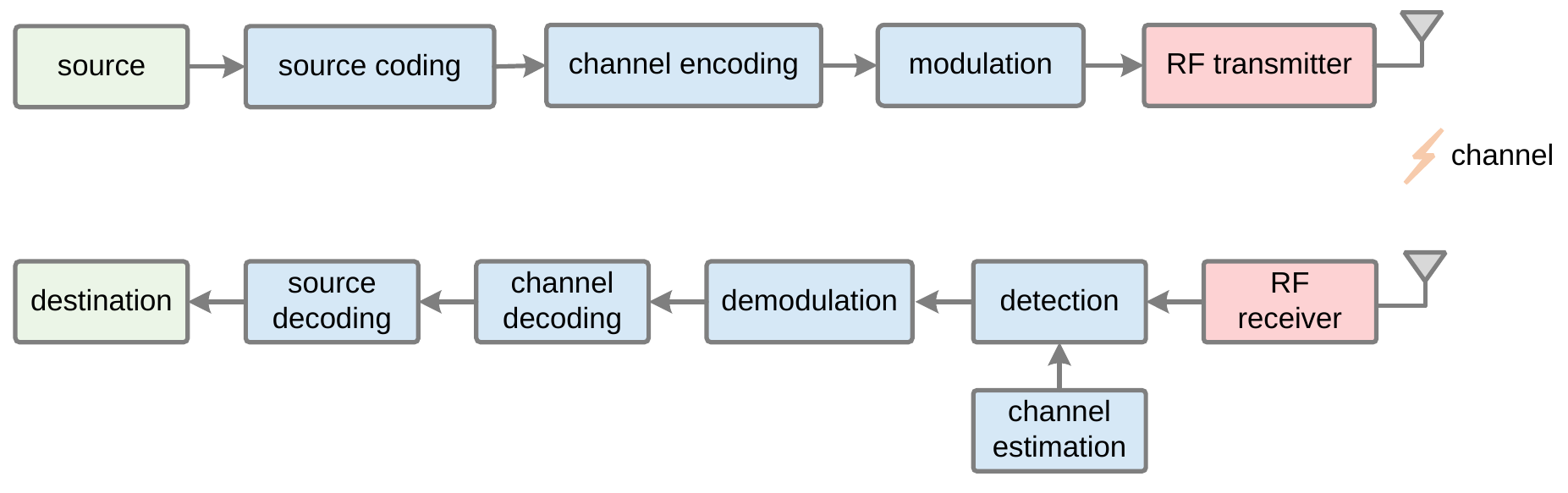}
		\centering\caption{A typical communication system diagram with blocks including source encoding/decoding, channel encoding/decoding, modulation/demodulation, channel estimation and detection, and RF transceiving. These signal processing blocks are optimized individually to achieve reliable communication between the source and target destination.}
		\label{communication_system_model}
	\end{figure*}
	
	The general classic communication system architecture is constructed as a block structure as shown in Fig. \ref{communication_system_model}, and multiple algorithms solidly founded on expert knowledge have been developed in long-term research to optimize each processing block therein. Previous studies have attempted to leverage conventional ML approaches, such as SVM and small feedforward NNs, as alternative algorithms for individual tasks. DL architectures have recently been introduced into several processing blocks to adapt to emerging complex communication scenarios or outperform conventional communication algorithms. This section presents some examples of DL applications that cover modulation recognition, channel decoding, and detection.

	\subsection{Modulation Recognition}
	\label{modulation recognition}
	Modulation recognition aims to distinguish modulation schemes of the received noisy signals, which is important to facilitate the communication among different communication systems, or interfere and monitor enemies for military use. Studies on modulation recognition have been conducted for many years using conventional algorithms that are divided into two categories, namely, decision-theoretic and pattern recognition approaches \cite{nandi1998algorithms}. These approaches have several common procedures, such as preprocessing, feature extraction, and classification. Previous studies have been keen to leverage ML algorithms (usually SVM and NN) due to their robustness, self-adaption, and nonlinear processing ability.
	
	\begin{figure}
		\centering
		\includegraphics[width=3.45in]{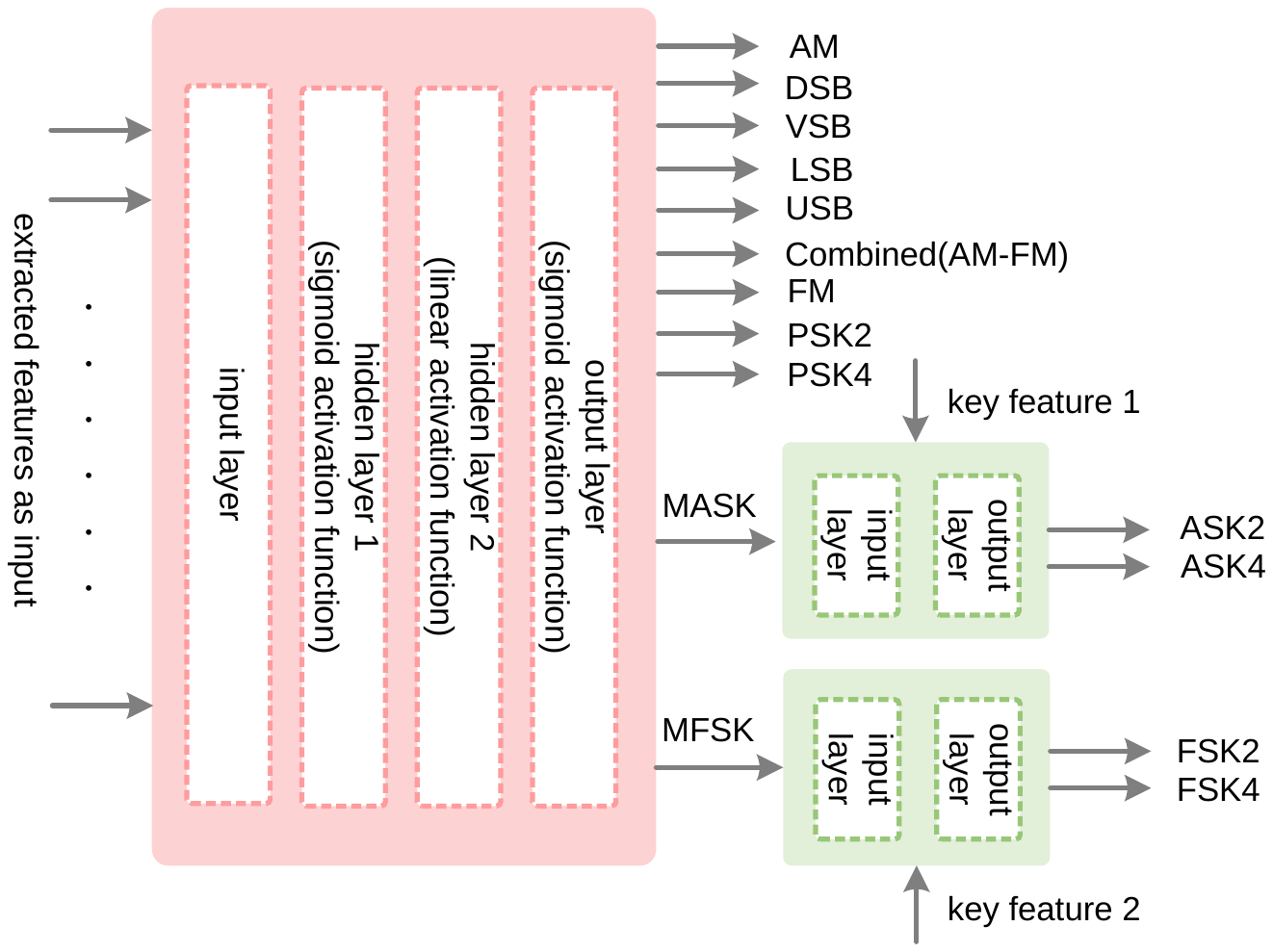}
		\centering\caption{An NN architecture for modulation recognition that consists of a four-layer NN and two two-layer NNs  \cite{nandi1998algorithms}. The former NN distinguishes most modulation schemes except for ASK and FSK. The latter NNs classify ASK2/ASK4 and FSK2/FSK4 with additional manually extracted key features.}
		\label{ANN for modulation}
	\end{figure}
	
	An NN architecture (Fig. \ref{ANN for modulation}) is proposed in \cite{nandi1998algorithms} as a powerful modulation classifier to discriminate noise-corrupted band-limited modulated signals from 13 types of digital modulation schemes (e.g., MPSK, MASK, and MFSK) and analog modulation schemes (e.g., AM, DSB, and FM). Similar to conventional expert feature analysis, this approach manually extracts features that characterize digital \cite{xin2004automatic} or analog modulations \cite{azzouz1996recognition} from the original signals, as well as the three fundamental parameters of instantaneous amplitude, phase, and frequency. A four-layer NN is then fed with these features to discriminate the modulation schemes, except the levels of MASK and MPSK, as identified by another two two-layer NNs. The abovementioned problem-solving procedures, such as manual feature extraction and NN classification, have been commonly applied in previous research (e.g., \cite{fehske2005new}). Their performance strongly depends on the extracted features due to the limited learning ability of conventional NNs.

	DL is known for its impressive learning capacity. Its introduction highlights the possibility of replacing artificially extracting features with automatic learning features from raw data to optimize end-to-end performance. For example, a CNN-based approach is proposed in \cite{O2017An} that learns single-carrier modulation schemes based on sampled raw time-series data in the radio domain. The CNN classifier is trained by 1.2M sequences for 128 complex-valued baseband IQ samples covering 10 different digital and analog modulation schemes that pass through a wireless channel with the effects of multipath fading, sample rate offset, and center frequency offset. The CNN-based modulation classifier dominates and outperforms two other approaches, namely, extreme gradient boosting with 1,000 estimators and a single scikit-learn tree working on the extracted expert features. The performance of the classifier improves along with increasing signal-to-noise ratio (SNR). However, such performance cannot be improved further at high SNR because the short-term nature of training samples confuses the CNN classifier between AM/FM if the underlying signal carries limited information and between QAM16/QAM64 that share constellation points. More samples must be used to eliminate such confusion and highlight the potential of the CNN architecture for further improvement.

	\subsection{Channel Decoding}
	\label{channel decoding}
	ML-based decoders have emerged in the 1990s \cite{gruber2017deep} because of the straightforward applications of NN to channel decoding. First, channel decoding algorithms focus on bit-level processing; therefore, bits, or the log-likelihood ratios (LLRs) of codewords, are conveniently treated as the inputs and expected outputs of NNs. Mainly in previous studies, the input and output nodes directly represent bits in codewords \cite{caid1990neural}, or use one-hot representation (i.e., each node represents one of all possible codewords) \cite{di1991use} such that the corresponding vector has only one element equal to 1 and other elements equal to 0. Second, unlike the difficulty in obtaining datasets in other scenarios, man-made codewords can generate sufficient training samples and achieve labeled outputs simultaneously. Furthermore, NN can learn from the noise version of codewords and avoid overfitting problems because the codewords are randomized to different samples by the channel noise in each training epoch.
	
	Compared with conventional decoders that are designed strictly based on information theory and often follow an iteration process that leads to high latency, the NN architecture does not require expert knowledge. After training a decoder, the decoding process becomes simple with low latency. Furthermore, the well-developed conventional decoding algorithms can serve as benchmarks for comparing the performance of newly proposed DL-based methods.
	
	Despite its advantages, the NN-based decoder is fundamentally restricted by its dimensionality (i.e., the training complexity increases exponentially along with block length) \cite{wang1996artificial} to learn fully from and classify codewords, thereby limiting its scalability. Fortunately, DL algorithms provide the potential to this problem. Aside from its inherent parallel implementation for complex computations, a multi-layer architecture with realizable training methods offers DL with an excellent learning capacity. Recent studies have leveraged these advantages to address the issue of dimensionality by introducing DL to well-developed iterative algorithms (i.e., unfolding the iterative structure to the layer structure) and by generalizing from limited codewords (i.e., inferring from seen codewords to unseen ones).
	
	The fully connected DNN-based decoder proposed in \cite{nachmani2016learning}, which falls under the first category, aims to improve the performance of the belief propagation (BP) algorithm in decoding high-density parity check (HDPC) codes. BP algorithm can achieve near Shannon capacity in decoding low-density parity check codes but struggles at decoding HDPC codes, such as Bose-Chaudhuri-Hocquenghem (BCH) codes, that are commonly used today.
	Conventional BP decoders can be constructed using a Tanner graph, where each variable node is connected to some check nodes. In each iteration, a variable (check) node transmits a message to one of its connected check (variable) nodes based on all messages received from the other connected check (variable) nodes. Thus, the BP algorithm with $L$ iterations can be unfolded as 2$L$ hidden layers in a fully connected DNN architecture, where each hidden layer contains the same number of neurons that represent the edges in the Tanner graph. These neurons will output the messages transmitted on corresponding edges. In other words, the odd (even) hidden layers output messages that are transmitted from the variable (check) nodes to the check (variable) nodes that are associated with the neurons or edges in the Tanner graph.
	
	The input and output are vectors of size $N$ that represent $N$-dimensional LLRs received from channels and $N$-bits decoded codewords, respectively. The equations for calculating the messages based on expert knowledge are then applied to the corresponding layers, and the final marginalization of the BP algorithm is achieved after the last layer (i.e., after the last iteration), which is involved in the loss function. The only difference of the DNN-based BP algorithm from the original algorithm is that weights are added to these equations or are assigned to the edges in the Tanner graph. Thus, the DNN-based decoder shares the same decoding structure of the Tanner graph; however, the messages are propagated on the weighted edges.
	
\begin{figure}
		\centering
		\includegraphics[width=3.55in]{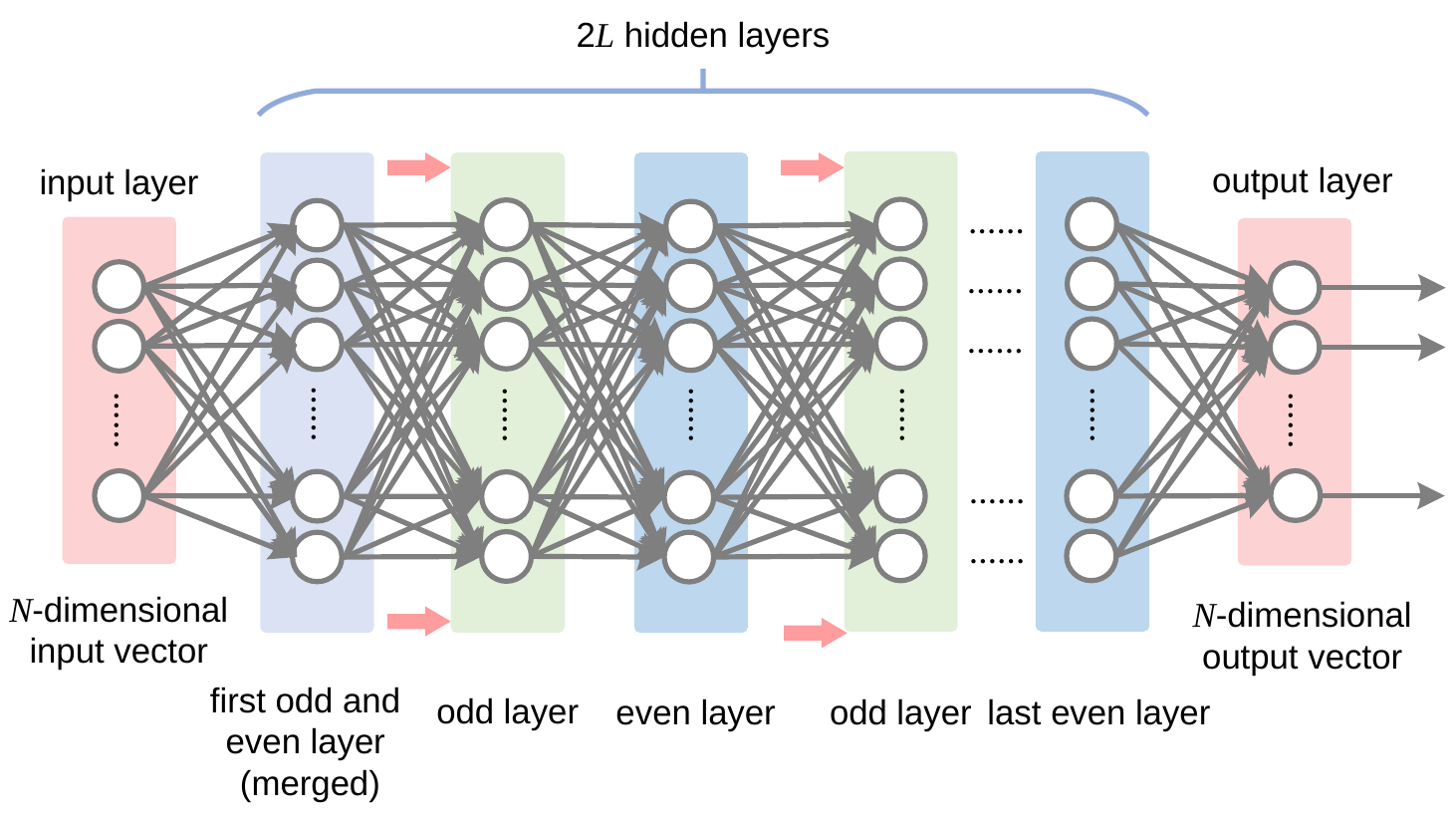}
		\centering\caption{A fully connected DNN-based BP decoder that unfolds conventional BP algorithms with $L$ iterations to $2L$ hidden layers \cite{nachmani2016learning}.}		
		\label{BP_linear}
	\end{figure}

	The DNN-based BP decoder proposed in \cite{nachmani2016learning} is shown in Fig. \ref{BP_linear}. The first two hidden layers are merged into one layer because the check nodes do not contain any information in the first iteration. LLR messages (i.e., inputs) are needed by the variable nodes to calculate the outgoing messages according to the corresponding BP formulas. These messages are represented as red arrows at the odd layers in Fig. \ref{BP_linear}. The DNN-based BP decoder preserves the property of BP with essentially similar structures and its performance is independent from the transmitted codewords. Thus, the network can be trained by the noise version of a single codeword (i.e., zero codeword) that belongs to all linear codes.
	
	\begin{figure}
		\centering
		\includegraphics[width=2.8in]{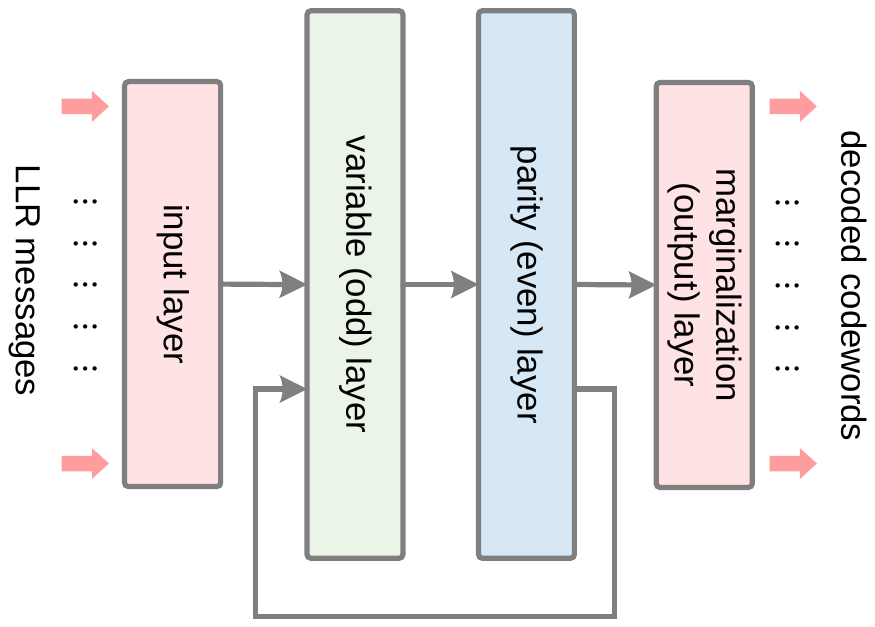}
		\centering\caption{A BP-RNN decoder architecture with a variable layer and a parity layer \cite{nachmani2017rnn}. The output of the parity layer is fed into the variable layer in the next time step, so that $L$ time steps represent $L$ iterations of the conventional BP algorithm.} 		
		\label{BP_RNN}
	\end{figure}
	
	After training the codewords that pass through the additive white Gaussian noise (AWGN) channel with SNRs ranging from 1 dB to 6 dB, a DNN-based BP decoder for BCH (15,11) with 10 hidden layers (i.e., 5 full iterations) achieves close to maximum likelihood results. The performance of the DNN-Based decoder degrades on large BCH codes; nevertheless, the decoder consistently outperforms the conventional BP algorithm. One interpretation given by \cite{nachmani2016learning} is that properly weighting the transmitted messages compensates for the small cycle effect in the original Tanner graph. Furthermore, the final marginalization information is considered obtainable after each odd hidden layer. Therefore, a modified architecture called multiloss, which adds such information into the loss function, is proposed to increase the gradient update and allow lower layers. The multiloss architecture achieves a better bit-error rate (BER) performance compared with the previous DNN-based BP decoder.
	
	In \cite{nachmani2017rnn}, the aforementioned fully connected DNN-based BP decoder is transformed into an RNN architecture, which is named as BP-RNN decoder, by unifying the weights in each iteration and feeding back the outputs of parity layers into the inputs of variable layers, as shown in Fig. \ref{BP_RNN}. The number of time steps equals to that of iterations. This process significantly reduces the number of parameters and results in a performance that is comparable with that of the former decoder. The multiloss concept is also adopted in this architecture. These proposed methods outperform the plain BP algorithm whether on the regular or sparser Tanner graph representations of the codes. A modified random redundant iterative decoding (mRRD)-RNN decoder, which combines the BP-RNN decoder with mRRD algorithms \cite{dimnik2009improved} by replacing its BP blocks, is proposed and outperforms the plain mRRD decoder with only a slight increase in complexity.

\begin{figure}
		\centering
		\includegraphics[width=3.5in]{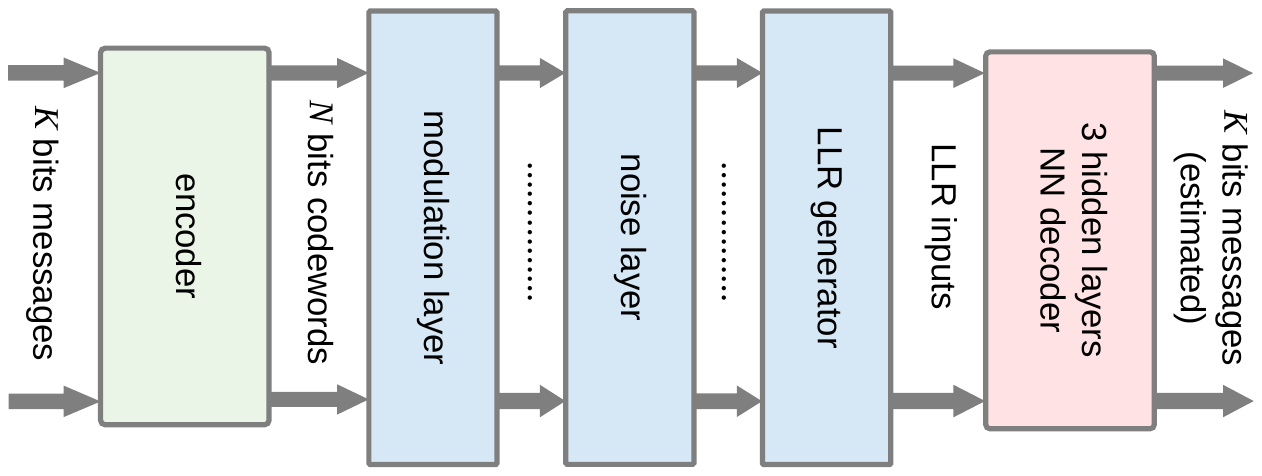}
		\centering\caption{A plain DNN architecture for channel decoding to decode $k$ bits messages from $N$ bits noisy codewords \cite{gruber2017deep}.}
		\label{DNN_decode}
	\end{figure}

	A plain DNN architecture called NN decoder (NND), is proposed in \cite{gruber2017deep} to decode codewords of length $N$ with  $K$ information bits, as shown in Fig. \ref{DNN_decode}. An encoded codeword of length $N$ passes through a modulation layer and an AWGN channel layer to represent the communication channel effect. The LLR information of this noisy codeword is then generated as an $N$-dimensional input of the network with three hidden layers that is trained to output $K$ estimated information bits. This NND works for 16-bit length random and structured codes (e.g., polar codes) and achieves maximum a posteriori (MAP) performance; however, such performance degrades along with an increasing number of information bits. Fortunately, the NND can generalize a subset of codewords (for training) to unseen codewords when decoding structured codes so that it is promising to address the issue of dimensionality by learning a form of decoding algorithm.
	
	Other advantages, such as parallel architecture and one-shot decoding (i.e., no iterations) with low latency, makes the NND a promising alternative channel-decoding algorithm. The authors in \cite{gruber2017deep} suggest the existence of an optimal SNR for training to classify the codewords over arbitrary SNRs, and argue that having more training epochs can lead to better performance. Training with direct channel values or LLR while using mean squared error or binary cross-entropy as a loss function has no significant effect on the final results.
	
	To further scale DL-based decoding to large codewords, several former NNDs, with each decoding a sub-codeword, are combined in \cite{cammerer2017scaling}. These NNDs are firstly trained individually to meet the MAP performance, and then combined to replace the sub-blocks of a conventional decoder for polar codes. Thus, a large codeword is concurrently decoded. Specifically, the encoding graph of the polar codes defined as partitionable codes in \cite{cammerer2017scaling} can be partitioned into sub-blocks that can be decoded independently. Thus, the corresponding BP decoder is partitioned into the sub-blocks replaced by NN architectures that are connected in the remaining BP decoding stages.
	
\begin{figure}
		\centering
		\includegraphics[width=2.7in]{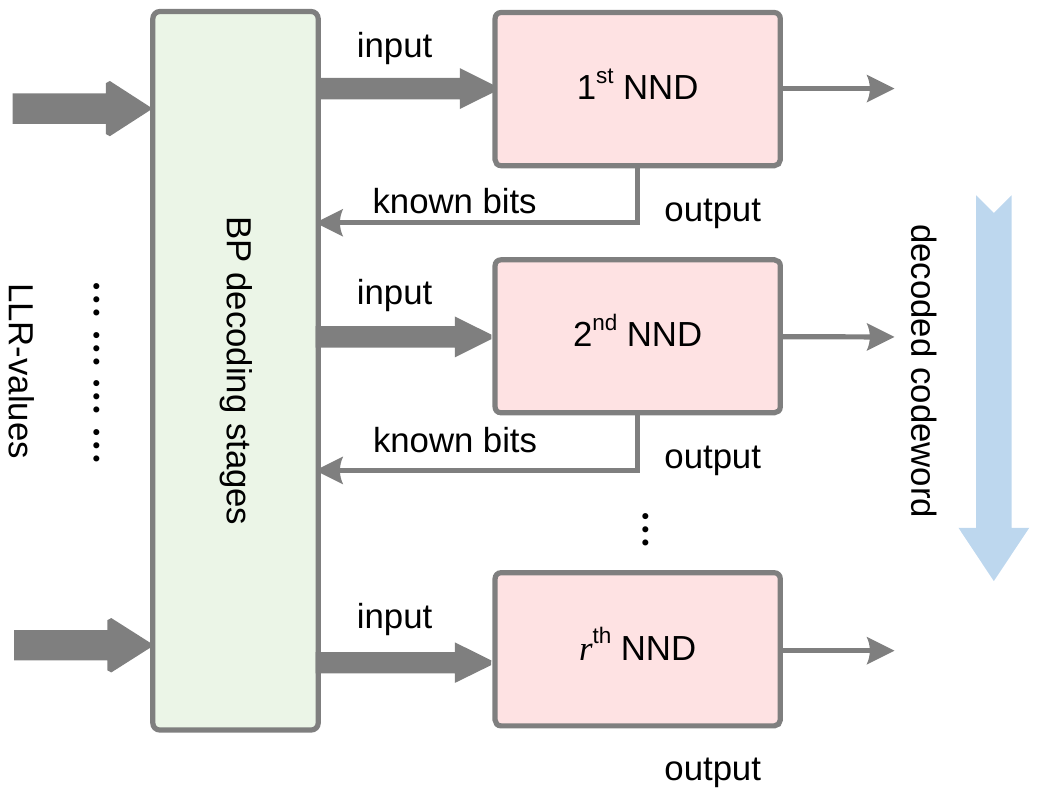}
		\centering\caption{A partitioned NN decoding architecture for polar codes with each NND decoding a sub-codeword \cite{cammerer2017scaling}.}
		\label{NN_partition}
	\end{figure}

	A partitioned neural network (PNN) architecture is proposed as shown in Fig. \ref{NN_partition}. The received LLR values from the channel are propagated step by step corresponding to the BP update algorithm and arrive at the first NND to decode the first sub-codeword, which is then propagated in the remaining BP decoding steps as known bits. This decoding process continues until all sub-codewords are sequentially decoded. Pipeline implementation can also be applied to decode multiple codewords. This proposed PNN architecture can compete well with conventional successive cancellation and BP decoding algorithms; however, its performance deteriorates along with an increasing number of partitioned sub-blocks, thereby limiting its application for large codes. Nevertheless, PNN still offers a promising solution to dimensionality problems.
	
	\subsection{Detection}
	\label{detection}
	
	Along with the increasing application of advanced communication systems with promising performance, capacity, and resources (e.g., massive MIMO and mmWave), the available communication scenarios or communication channels are becoming increasingly complex, thereby increasing the computational complexity of channel models and the corresponding detection algorithms.  Conventional (iterative) detection algorithms form a computational bottleneck in real-time implementation.
	In comparison, given their expressive capacity and tractable parameter optimization, DL methods can be used for detection by unfolding specific iterative detection algorithms (similar to channel decoding) and for making a tradeoff between accuracy and complexity by leveraging flexible layer structures. Data detection becomes a simple forward pass through the network and can be performed in real time.
	
	\begin{figure}
		\centering
		\includegraphics[width=3.25in]{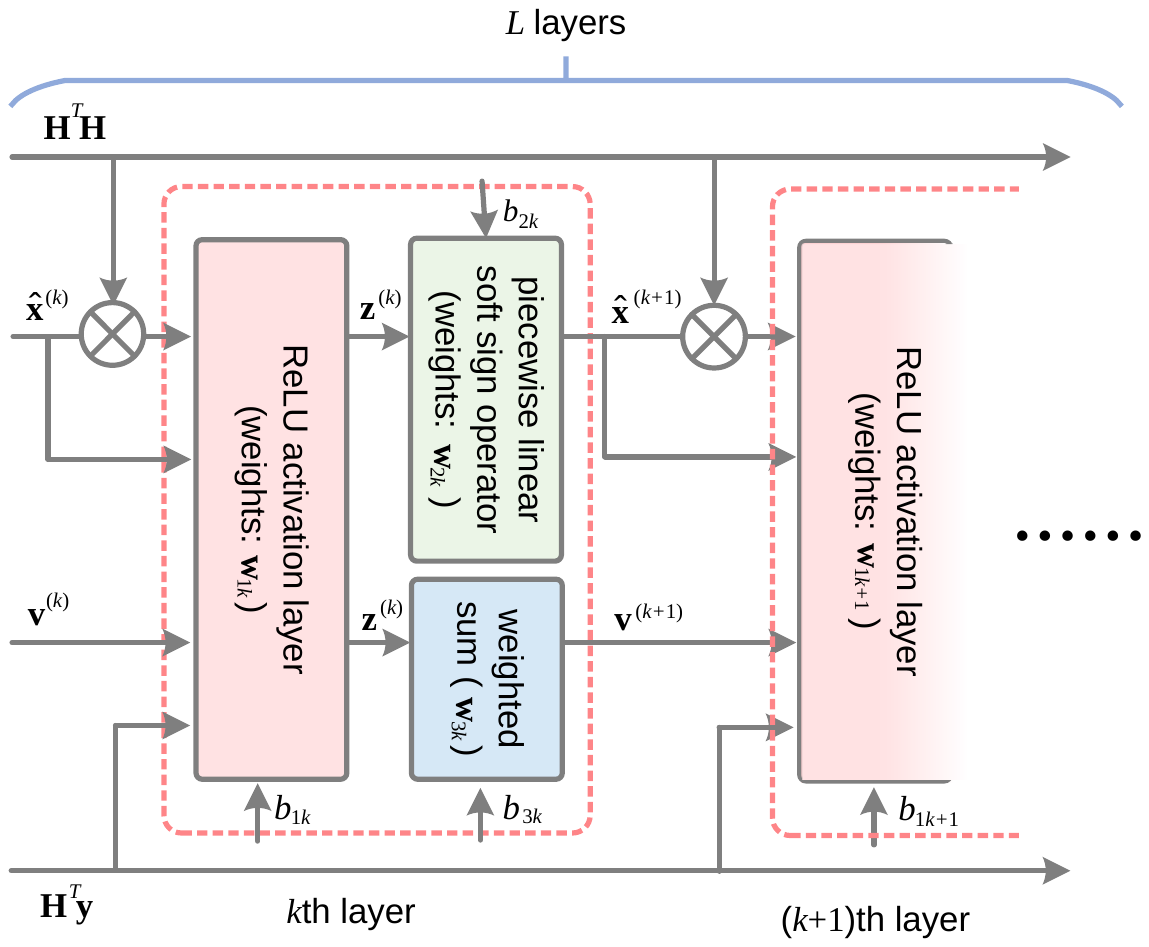}
		\centering\caption{A single layer structure of the DetNet architecture \cite{samuel2017deep}. $L$ layers of the DetNet represent $L$ iterations of the projected gradient descent algorithm.}		
		\label{DetNet}
	\end{figure}
	
	In \cite{samuel2017deep}, a DL-based detector called DetNet, which aims to reconstruct the transmitted $\bf x$ by treating the received $\bf y$ and channel matrix $ \bf H $ (assumed to be perfectly known) as inputs, is introduced by unfolding a projected gradient descent algorithm for maximum likelihood optimization, in which the iteration is computed as follows given by \cite{samuel2017deep}:
	\begin{equation}
		\begin{aligned}
			{\hat{\bf x}}^{(k+1)}
			&=\Pi{\left(\hat{\bf x}^{(k)}-\delta_{k}\dfrac{\partial\Vert \bf y-\bf H \bf x\Vert^{2}}{\partial \bf x}\bigg|_{{\bf x=\hat{ x}}^{(k)}} \right)}\\
			&=\Pi{\left(\hat{\bf x}^{(k)}-\delta_{k}{{\bf {H}}^{T} \bf y}+\delta_{k} {\bf H}^{T}{\bf H}{\bf x}^{(k)} \right)},
			\label{projected gradient descend}
		\end{aligned}
	\end{equation}
	where $\delta_{k}$ is the step size, $\hat{\bf x}^{(k)}$ is the estimate in the $k$th iteration, and
	$\Pi(\cdot)$ is a nonlinear projection operating on a linear combination of factors that is formed by the data. These data are lifted into the higher dimension and operated by ReLU activation function (represented by $\rho$ in (\ref{for layer}) given by \cite{samuel2017deep}) to leverage the DNN architecture. As shown in Fig. \ref{DetNet}, each original iteration is unfolded to the following layer:
	\begin{equation}
		\begin{aligned}
			&{\bf z}^{(k)} = \rho \left({\bf W}_{1k}\begin{bmatrix} {\bf H}^{T} \bf y \\\hat{\bf x}^{(k)} \\ {\bf H}^{T}{\bf H} \hat{\bf x}^{(k)} \\ {\bf v}^{(k)}\end{bmatrix}\right)+{\bf b}_{1k},\\
			&{\bf \hat{x}}^{(k+1)}  =\psi_{t_{k}}{\left({\bf W}_{2k} {\bf z}^{(k)}+{\bf b}_{2k} \right)}, \\
			&{\bf \hat{v}}^{(k+1)} ={\bf W}_{3k} {\bf z}^{(k)}+{\bf b}_{3k},\\
			&\hat{\bf x}^{(1)} = {\bf 0},
			\label{for layer}
		\end{aligned}
	\end{equation}
	where $\{ {\bf W}_{1k},{\bf W}_{2k},{\bf W}_{3k} \}$ and $\{ {\bf b}_{1k},{\bf b}_{2k},{\bf b}_{3k} \}$ are the weights and bias, respectively, and  $\psi_{t}(\cdot)$ is a defined piecewise linear soft sign operator. $L$ layers (i.e., $L$ iterations) are applied in DetNet. A loss function covering the outputs of all layers is adopted to prevent the gradients from vanishing, and the Adam optimization algorithm is used for the training.
	
	To test the robust performance of DetNet in complex channels and generalization, two channel scenarios are considered, namely, the fixed channel (FC) model with a deterministic yet ill condition and the varying channel (VC) model with an $\bf H$ that is randomly generated by a known distribution. Compared with the conventional approximate message passing (AMP) and semidefinite relaxation (SDR) algorithms that provide near-optimal detection accuracy, the simulation results in the FC scenario indicate that DetNet outperforms AMP  and achieves similar accuracy as SDR but runs 30 times faster, whether trained by the FC or VC channel. Therefore, DetNet remains robust in ill-conditioned channels and is generalized during training to detect arbitrary channels. Similarly, DetNet works 30 times faster than SDR in the VC channel but shows a comparable performance. DetNet works even faster at shallow layers but with less accuracy, thereby illustrating a tradeoff between accuracy and complexity. Accordingly, the DetNet presents a promising solution to computationally challenging detection tasks in advanced complex scenarios.
	
	Despite the complex channel models, no mathematically tractable channel models are available to characterize the physical propagation process accurately in highly complex cases, such as molecular and underwater communications. Therefore, new detection approaches that do not require channel information must be devised for the novel systems. DL offers a promising solution to this problem because of its expressive capacity, data-driven characteristic, non-requirement for a defined channel model, and ability to optimize end-to-end performance. In \cite{farsad2017detection} a fully connected DNN, CNN, and RNN are applied for detection in a molecular communication system. A molecular communication experimental platform is established to generate an adequate dataset by repeatedly transmitting a consecutive sequence of  $N$ symbols from $M$ possible types. Chemical signals, acids (representing bit-0), and bases (representing bit-1) are used to encode pH level information. After transmission in water, the pH values in each symbol interval are treated as the received signals. In a simple baseline detection, bit-0 or bit-1 is detected in each symbol interval according to a decrease or increase in pH values, respectively, which is decided by the difference between the subintervals that the original symbol interval is divided into (i.e., positive for bit-1 and negative for bit-0).
	
\begin{figure}
		\centering
		\includegraphics[width=3.5in]{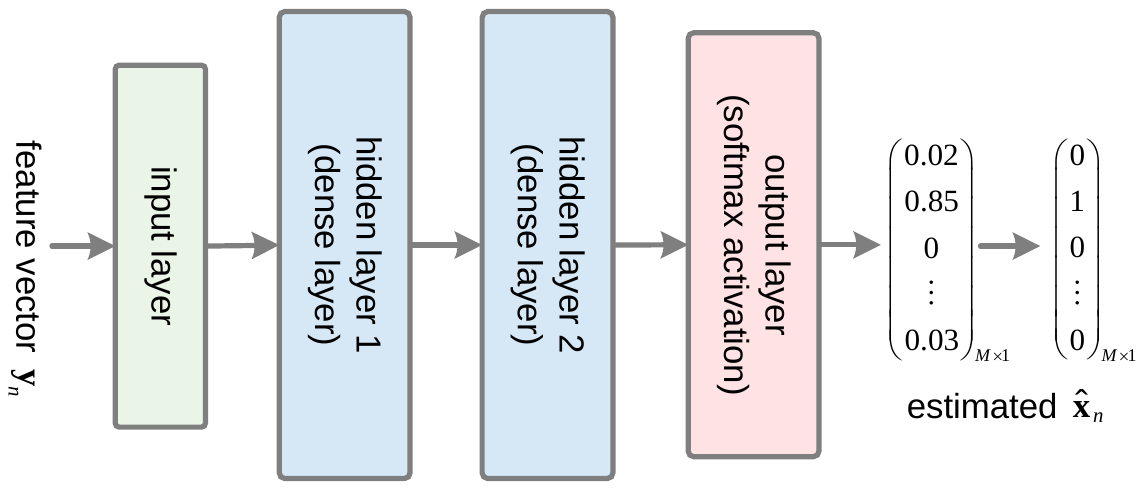}
		\centering\caption{A dense-Net for symbol-to-symbol detection to detect an estimated $\hat{\bf x}_{n}$ in one-hot representation \cite{farsad2017detection}.}		
		\label{Dense_Net}
	\end{figure}

\begin{figure}
		\centering
		\includegraphics[width=2.2in]{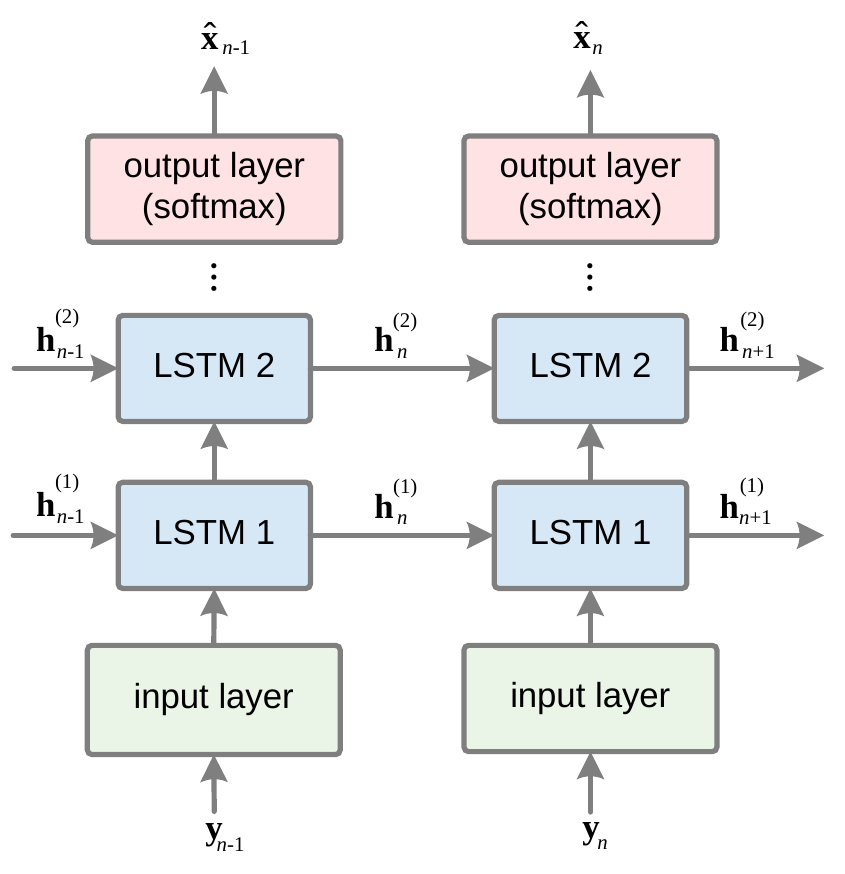}
		\centering\caption{A LSTM-based detector for sequence detection with $L$ LSTM layers followed by the dense output layer with softmax activation function \cite{farsad2017detection}.}		
		\label{RNN_Net}
	\end{figure}

	When designing the DL-based detector, a simple memoryless system is considered such that the  $n$th received signal is determined by the $n$th transmitted signal ${\bf x}_{n}$. Symbol-to-symbol detection can be implemented using Dense-Net, a basic fully connected DNN architecture (Fig. \ref{Dense_Net}) that inputs the former pH difference values and some absolute values representing as the received feature vector ${\bf y}_{n}$ and outputs an $M$-dimensional probability vector. A CNN-based detector is also introduced to adapt to the effect of random shift. The system with inter symbol interference (ISI) and memory presents a highly sophisticated yet realistic scenario that requires sequence detection and can be implemented by the LSTM network, a typical algorithm for sequence processing belonging to RNN. As shown in Fig. \ref{RNN_Net}, the $n$th estimated $\hat{\bf x}_{n}$ is detected from the previously and currently received signals.
	
	The simulation results demonstrate that all these DL-based detectors outperform the baseline, whereas the LSTM-based detector shows an outstanding performance in the molecular communication system with ISI. This result validates the potential application of DL in future novel systems and highlights the importance of selecting suitable DL architectures that adequately reflect the characteristics of the physical channel.
	
	Aside from complex scenarios, DL can be applied to well-researched channel conditions to enhance its performance further. For example, in \cite{ye2017power}, a five-layer fully connected DNN is embedded into an OFDM system for channel estimation and detection by treating the channel as a black box. During the offline training, the original data and pilots are formed into frames, and pass through a statistic channel model with distortion after inverse discrete Fourier transform processing and adding a cyclic prefix (CP), so that the received signal in time domain is generated. The frequency domain complex signal is obtained by removing CP from the former time domain signal and performing discrete Fourier transform. The frequency domain signal comprising data and pilot information, is then fed into the DNN detector to reconstruct the transmitted data in frequency domain. In comparison with the conventional minimum mean square error (MMSE) method, the DNN detector achieves comparable performance in online testing but shows better performance when less pilots or no CP is used or when clipping distortion is introduced to reduce the peak-to-average power ratio. The DNN detector also shows a stable performance when tested in channel models with different delays and path numbers, thereby highlighting its robustness and ability to further improve conventional communication systems.
	
	\section{DL as a Novel Communication Architecture}
	\label{application autoencoder}
	Despite their promising performance, DL-based algorithms are often proposed as alternatives for one or two processing blocks of the classic block-structure communication system (Fig. \ref{communication_system_model}). However, basic communication tasks aim to propagate signals from one point to another through a physical communication channel, and use a transmitter and receiver to manage the practical channel effect and ensure reliability when no rigid block structure is requested. Thus, the optimization in each block cannot guarantee global optimization for the communication problem because performance improvements can be achieved if two or more blocks are jointly optimized. O'Shea et al. recast communication as an end-to-end reconstruction optimization task and propose a novel concept based on DL theories to represent the simplified system as an autoencoder. They first introduce an autoencoder to the field of communication \cite{O2017An} and then propose radio transformer network (RTN) to combine the DL architecture with the knowledge of communication experts. This autoencoder system is also extended to multi-user and MIMO scenarios in \cite{O2017An} and \cite{o2017deep}, respectively.
	\begin{figure}
		\centering
		\includegraphics[width=3.0in]{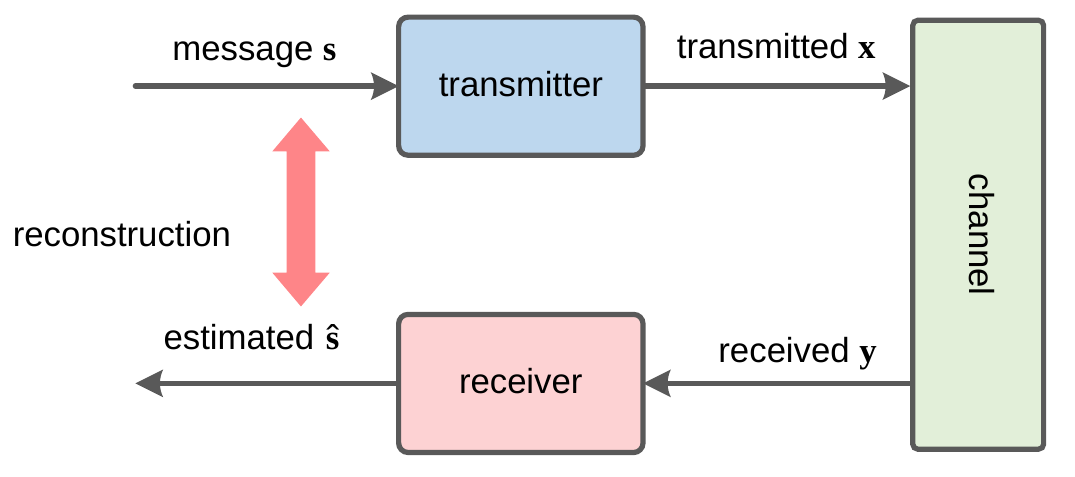}
		\centering\caption{A simple form of communication system that reconstructs the transmitted message at the receiver side \cite{dorner2017deep}.}		
		\label{simple communication system}
	\end{figure}
	
	\subsection{Autoencoder-based End-to-end System}
	\label{end-to-end}
	In \cite{O2017An}, communication is considered an end-to-end reconstruction problem where the transmitted messages are reconstructed at the receiver side over a physical channel (Fig. \ref{simple communication system}). Therefore, the autoencoder can represent the entire communication system and jointly optimize the transmitter and receiver over an AWGN channel. An original autoencoder is an unsupervised DL algorithm that learns a compressed representation form of inputs that can be used to reconstruct the inputs at the output layer. In the proposed approach, the transmitter and receiver are represented as fully connected DNNs, whereas the AWGN channel between them is represented as a simple noise layer with a certain variance. Thus, the communication system can be regarded as a large autoencoder that aims to learn from $\bf s$, which is one out of the  $M$ possible messages for propagation, to generate a representation of the transmitted signal $\bf x$ that is robust against the imperfect channel. Therefore, at the receiver side, the original message can be reconstructed as $\hat{\bf s}$ with a low error rate by learning from the received $\bf y$. The entire autoencoder-based communication system is trained to achieve end-to-end performance, such as BER or block error rate (BLER). However, this process may add redundancy in representation $\bf x$, which differs from the original DL autoencoder that learns to compress inputs restoratively.
	
\begin{figure}
		\centering
		\includegraphics[width=3.5in]{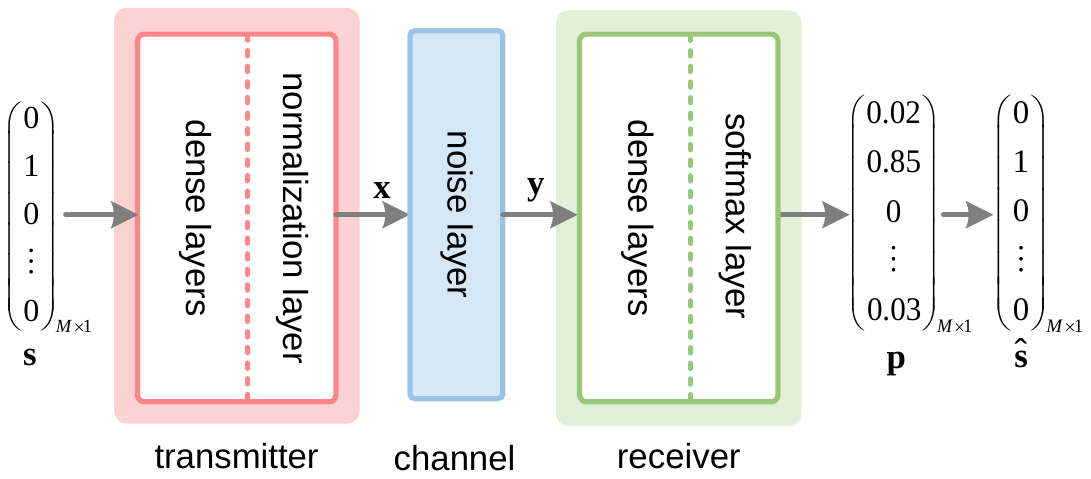}
		\centering\caption{A simple autoencoder for an end-to-end communication system \cite{O2017An} that encodes $\bf s$ in one hot representation to an $N$-dimensional transmitted signal $\bf x$. This encoded signal after adding a noise $\bf y$ is then decoded to an $M$-dimensional probability vector $\bf p$, and then $\hat{\bf {s}}$ is determined.}		
		\label{autoencoder}
	\end{figure}

	In an implementation example (Fig. \ref{autoencoder}) \cite{O2017An}, $\bf s$ is represented as an \textit{M}-dimensional one-hot vector. Therefore, $K={\rm log}_{2}(M) $ bits are transmitted simultaneously. After being fed into the DNN transmitter with multiple dense layers followed by a normalization layer, an $N$-dimensional vector $\bf x$ with energy constraints is generated. Therefore, the communication rate of such system is $ R=K/N $. The received $N$-dimensional signal $\bf y$ noised by a channel represented as a conditional probability density function $ p(\bf y|\bf x) $ is subsequently learned by the DNN receiver with multiple dense layers. The last layer of the receiver is a softmax activation layer that outputs an $M$-dimensional probability vector $\bf p$, in which the sum of its elements ($ \geq 0 $) is equal to 1. The index of the largest element with the highest probability determines which of the $M$ possible messages is the decoded $\hat{\bf s}$. The autoencoder is trained by SGD at a fixed SNR with categorical cross-entropy as a loss function to optimize BLER performance. The autoencoder-based communication system achieves a comparable or better performance than the conventional BPSK with Hamming code, thereby indicating that this system has learned a joint coding and modulation scheme.
	
	Such autoencoder architecture can solve other communication problems in the physical layer, such as pulse shaping and offset compensation, by dealing with IQ samples and can be applied to complex scenarios where communication channels are unknown.
	\subsection{Extended Architecture with Expert Knowledge}
	Aside from interpreting the communication system as a plain DL model, it is reasonable to consider introducing communication expert knowledge or adjusting the DL architecture to accommodate certain communication scenarios or accelerate the training phase. As shown in \cite{O2017An}, certain parametric transformations correspond to the channel effect. The inverse forms of these transformations can compensate for the channel distortion whereas the estimation of their parameters are highly related to received signals, thereby enabling the integration of communication knowledge into a DL system by generating parameters from these signals for such deterministic transformations.
	
\begin{figure}
		\centering
		\includegraphics[width=3.5in]{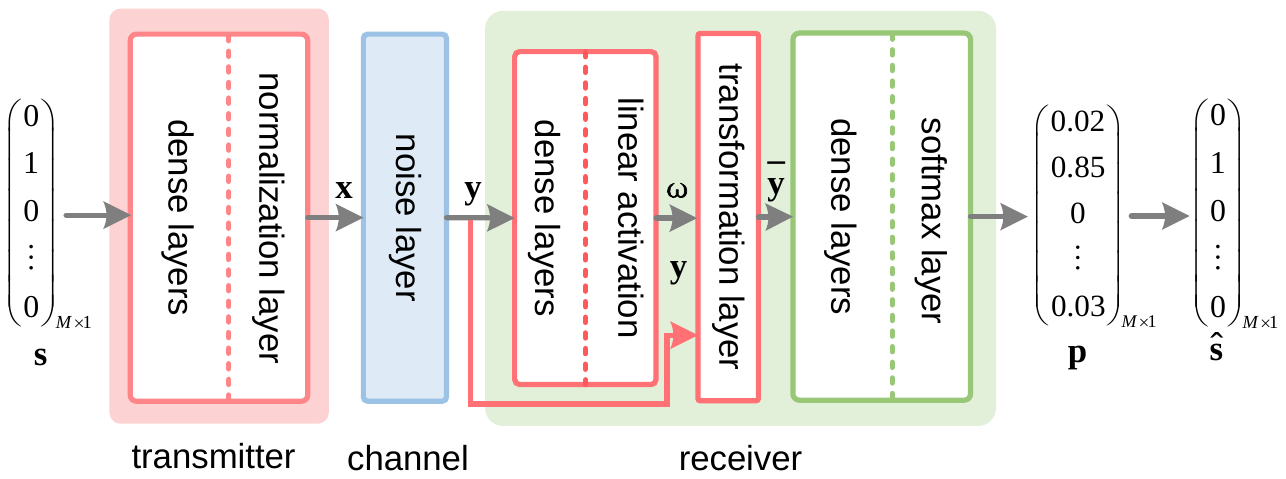}
		\centering\caption{A receiver implemented as an RTN \cite{O2017An}. In the receiver, a new block consisting of dense layers with linear activation function and a deterministic transformation layer, is added to the original autoencoder-based communication system.}		
		\label{RTN}
	\end{figure}

	RTN that extends the former DNN receiver by adding a parameter estimation module before learning to discriminate is proposed \cite{O2017An} (Fig. \ref{RTN}). Specifically, a parameter vector $ \bm \omega $ is learned from the received $\bf y$ by a fully connected DNN with linear activation in the last layer and then fed into a deterministic transformation layer that is parameterized by $ \bm \omega $, which corresponds to specific communication properties. The transformation is performed on $\bf y$ to generate a canonicalized $\overline{\bf y}$. The formerly learned DNN in the receiver with softmax activation completes the following discrimination task to output an estimated $ \hat{\bf s}$. In addition to its application to the receiver, the RTN architecture can be used for any scenario where deterministic transformations with estimated parameters exist. The simulation results indicate that the autoencoder with RTN outperforms and converges faster than the plain autoencoder architecture, thereby validating the effectiveness of combining the DL model with prior expert knowledge.
	
	\subsection{Autoencoder for Multi-user}
	\label{two user}
	
	\begin{figure}
		\centering
		\includegraphics[width=3.1in]{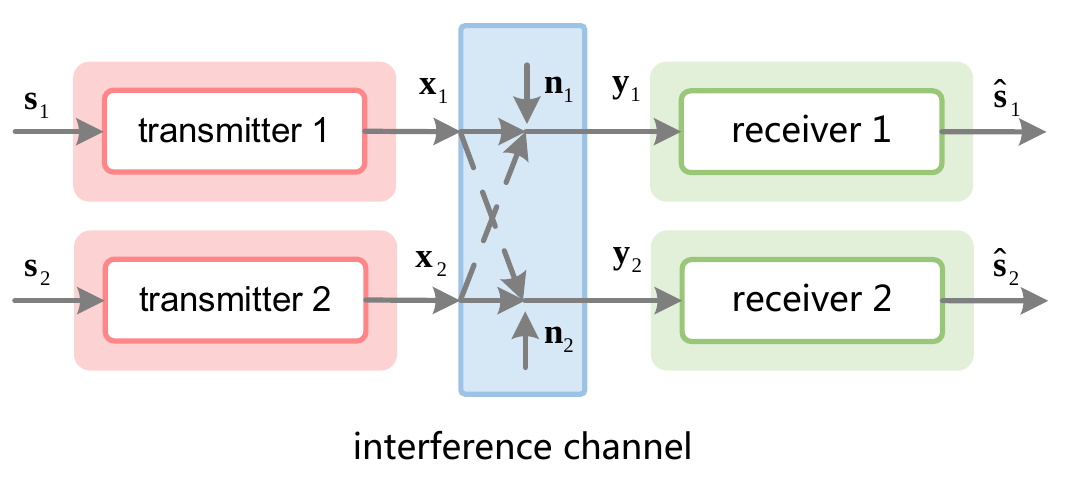}
		\centering\caption{Two-user scenario with interfering channel \cite{O2017An}. Each receiver has to detect their own messages based on received signals from two transmitters.}		
		\label{interfere_channel}
	\end{figure}
	
	The autoencoder system must be realizable for highly complex scenarios, such as multi-user communication over interference channels, to become a universal communication architecture. The application of autoencoders to a simple two-user scenario is explored in \cite{O2017An}, where two autoencoder-based transmitter-receiver pairs attempt to communicate simultaneously over the same interfering channel. The only difference of this scenario from the single-user case is that the entire system is trained to achieve conflicting goals with interference at the receiver side (Fig. \ref{interfere_channel}), that is, each transmitter-receiver pair aims to optimize the system to propagate their own messages accurately.

	One proposed method is to optimize the weighted sum of cross-entropy loss functions as $ J=\alpha J_{1}+(1-\alpha) J_{2} $. $J_{1}$ and $J_{2}$ represent the losses of the first and second transmitter-receiver pairs, respectively. $\alpha$ is a dynamic weighted factor ranging from 0 to 1 and is related to mini-batch parameters. The autoencoder system achieves the same or even better BLER performance at the same communication rate than conventional uncoded QAM schemes, thereby validating its potential application in multi-user cases.

	\subsection{Autoencoder for MIMO}
	In \cite{o2017deep}, the autoencoder communication system is extended to MIMO channels. Two types of conventional MIMO communication schemes are considered, namely, an open-loop system without channel state information (CSI) feedback and a closed-loop system with CSI feedback. Unlike the AWGN channel model used before, an { $r\times t$} MIMO channel response $\bf H$ is randomly generated before adding noise in the channel network block.
	
\begin{figure}
		\centering
		\includegraphics[width=3.5in]{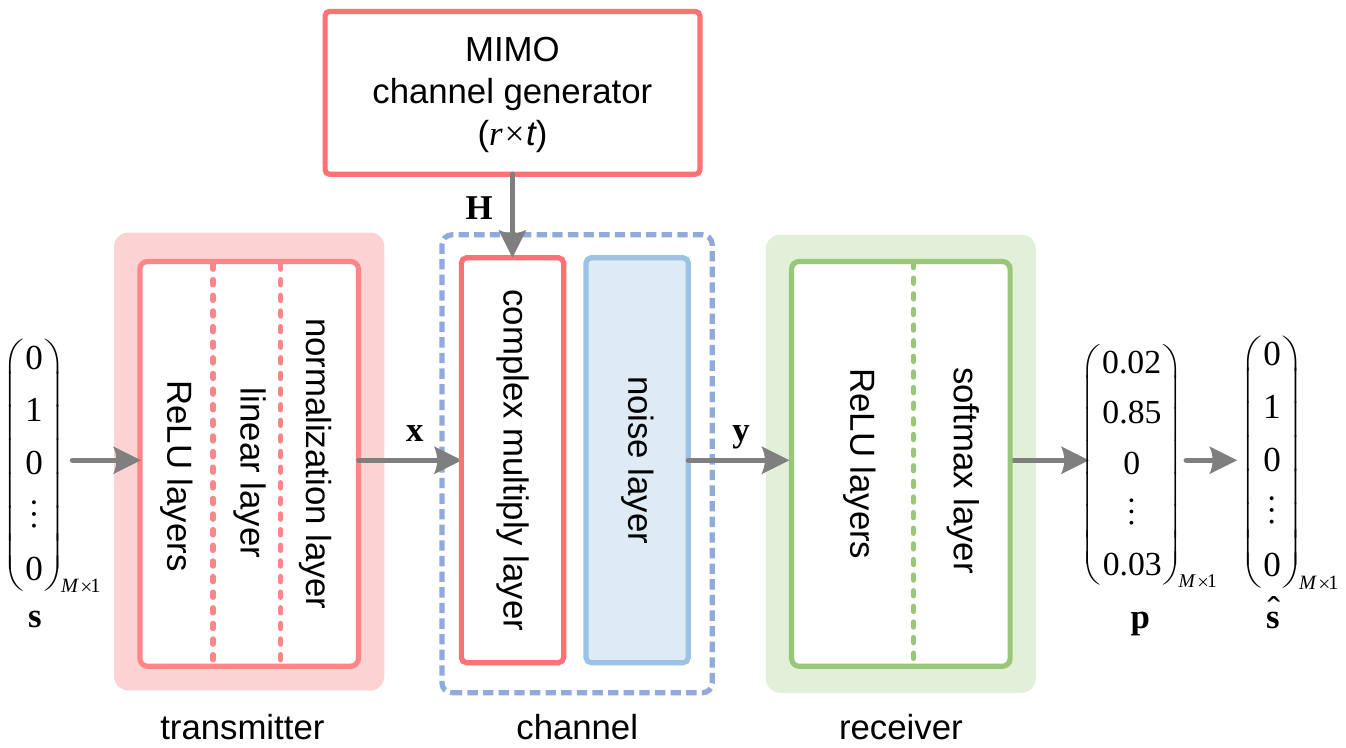}
		\centering\caption{A general MIMO channel autoencoder architecture \cite{o2017deep}. This system deals with the open-loop case where no CSI feedback exists.}		
		\label{no CSI}
	\end{figure}

	In the first open-loop case, the transmitter in the primary autoencoder system is modified to encode message ${\bf s}$ to the transmitted signal ${\bf x}$ as $t$ parallel streams of $N$ time samples. The message is then multiplied by a pre-defined $\bf H$ before passing through the noise layer to generate the received signal $\bf y$ as $r$ streams of $N$ time samples (Fig. \ref{no CSI}). Therefore, the estimated $\hat{\bf s}$ is learned from $\bf y$ in the following procedures. In the simulation of a $2 \times 1$ MIMO system, the adjusted autoencoder architecture outperforms the conventional open-loop MIMO schemes, such as Alamouti STBC with an MMSE receiver, when SNR $\geq15$dB.
	
\begin{figure}
		\centering
		\includegraphics[width=3.5in]{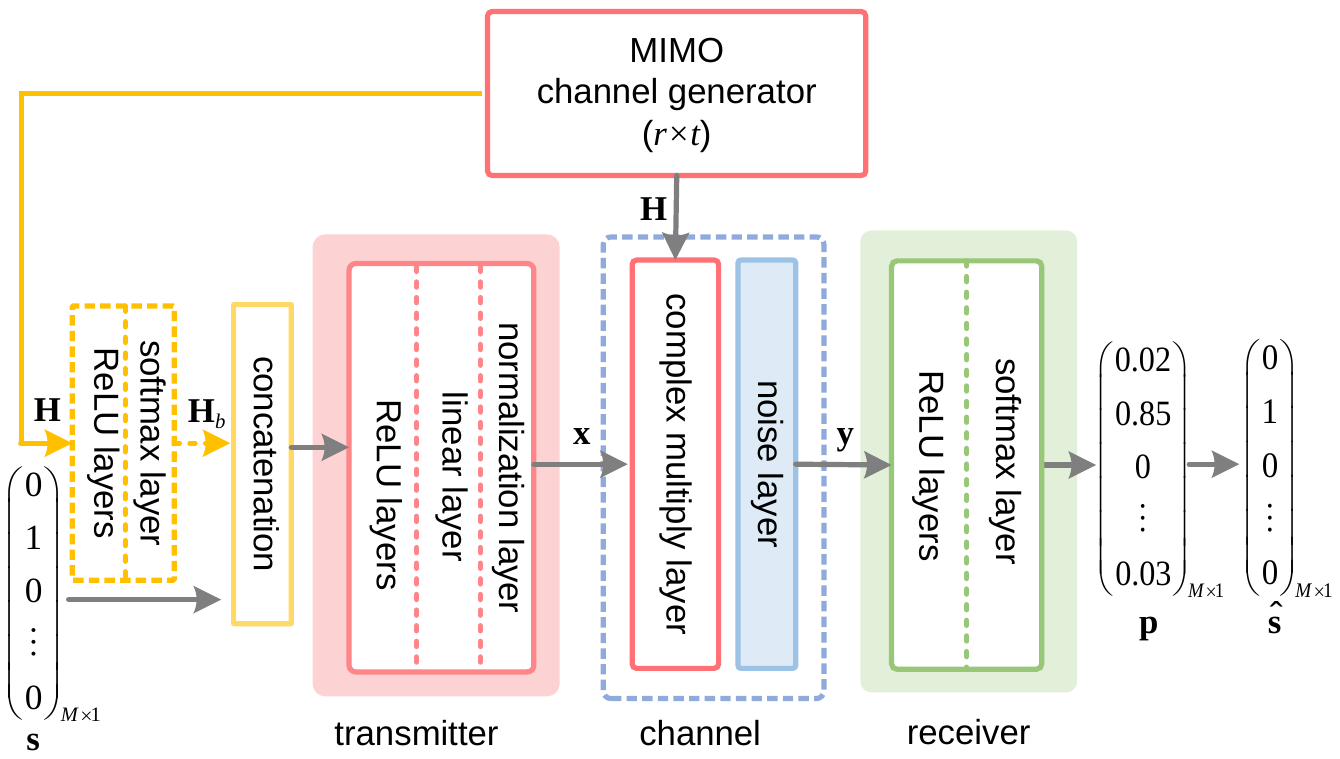}
		\centering\caption{A MIMO channel autoencoder for closed-loop scenarios with perfect CSI or quantized CSI feedback \cite{o2017deep}. In the perfect CSI case, a concatenation layer is added to the general MIMO autoencoder to concatenate $\bf H$ with $\bf s$ as inputs. In the quantized CSI case, a block in yellow dashed box to compact $\bf H$ as ${\bf H}_{b}$ before concatenation, is further added to the former architecture.} 		
		\label{perfect CSI}
	\end{figure}

	In the closed-loop case, an idealized situation is considered where the transmitter can obtain perfect CSI information from the receiver. Compared with the open-loop case, a feedback is added to the general MIMO autoencoder architecture, as shown in Fig. \ref{perfect CSI}. More specifically, The channel response $\bf H$ generated in the channel module is propagated to the transmitter as an input, which is concatenated with $\bf s$ before encoded to $\bf x$. The MIMO autoencoder with perfect CSI outperforms the conventional singular value decomposition-based precoding scheme at most SNRs.
	
	However, CSI errors exist in practice, such as inaccurate estimation, compaction, and unreliable feedback propagation. Therefore, a quantized-CSI scheme is introduced to the autoencoder to represent the practical situation. The only difference of this scheme from the former MIMO autoencoder for a perfect CSI is that before it is concatenated with message $\bf s$ at the transmitter side, the real-valued $\bf H$ is compacted as a $b$-bit vector ${\bf H}_{b}$ by another NN to represent $2^{b}$ modes, which is shown as the yellow dashed box in Fig. \ref{perfect CSI}. The simulation of a $2 \times 2$ system demonstrates that in contrast to the performance decrease of the conventional LLoyd algorithm with fewer bits of CSI, the autoencoder outperforms the perfect CSI in some quantized CSI cases. Thus, such quantization method helps the system achieve better convergence for each channel mode.

	\section{Discussion and Future Works}
	\label{future}
	The application of DL to the physical layer of wireless communication systems presents a new research field that is still in its early stage. Although previous studies have shown promising results, some extensive challenges are worth exploring further in future investigations.
	
	\subsubsection{Further extensions on extant researches}
	Emerging researches have attempted to introduce DL as an alternative for certain modules of the conventional communication system. The performance achieved in these researches inspire an extension of the proposed approaches to further applications or evolution.
	
	For example, the success of DL-based modulation recognition methods verifies the capability of DL in feature extraction and recognition. It implies the potential to apply DL for efficient recognition of other system parameters, such as recognizing source coding or channel coding schemes, extracting CSI, and learning characteristics from signals, which enables the wireless communication system more ``knowledgeable'' or ``intelligent'' in the physical layer. Combining the intelligence in the physical layer and upper layers straightforwardly, the communication systems can adapt to the features of transmitted signals and propagation environment automatically and achieve flexible deployment.
	
	In \cite{dorner2017deep}, the learned architecture has difficulties with VCs. Inspired by the competitive performance achieved by CSI feedback in the MIMO autoencoder system in \cite{o2017deep}, this condition can be improved if the channel information is sent to the receiver, as suggested by the expert knowledge from the communication domain. Similarly, the performance of the DL approaches presented in Figs. \ref{Dense_Net}, \ref{RNN_Net}, and \ref{autoencoder} can be further improved by introducing channel information.
	
	\subsubsection{Specialized DL architecture for communication}
	The DL architecture used for learning has an important influence on the final performance of the network, whereas the current design schemes for communication networks are simple. For certain algorithms with iterations (e.g., data detection \cite{samuel2017deep}, decoding \cite{nachmani2016learning}, and compressive sensing \cite{Mousavi2015CS,Mousavi2017CS,Lohit2017CS}), a straightforward method is to unfold the iterations as layer structures, and weights can be added to the original iterative formulas for training. The sub-blocks of a partitionable algorithm can also be replaced, as shown in \cite{cammerer2017scaling}. Despite their low complexity and comparable performance, these approaches only modify the conventional algorithms and do not create any significant differences or learn any new algorithms. Some studies \cite{gruber2017deep,O2017An} have applied plain DL methods to communication cases. However, these methods suffer from dimensionality problems because the complexity of networks and training phases grows exponentially along with the increasing number of messages or codewords, thereby limiting their application in practical scenarios.
	
	Therefore, expert knowledge from the communication domain must be introduced to DL architecture design. As presented in \cite{farsad2017detection}, the performance of the system can be improved using suitable DL networks that better characterize the channel conditions. The RTN proposed in \cite{O2017An}, represents the first attempt to add a priori knowledge to the plain DL architectures, and illustrates the benefits of reducing complexity and accelerating convergence. Despite its limited salability, RTN has inspired a design of specialized DL systems for communication fields based on the basic DL architecture. However, the specialized design is not limited to the proposed RTN architecture and worth extensive exploration. Novel and advanced DL-based systems that correspond to expert knowledge in propagation must be proposed to devise effective algorithms for future communication systems and address the limitations in their scalability.
	
	\subsubsection{Learning strategies and performance analysis}
	Although the recently proposed DL-based communication algorithms demonstrate a competitive performance, they lack solid foundations for theoretical derivation and analysis. The performance boundaries and the minimum dataset required for training also lack any certification. Moreover, given that the research on the application of DL to the physical layer of wireless communication systems is still in its early stage, the rules of learning strategies remain unknown and warrant further exploration. Unlike CV where the dataset is often represented as pixel values, the system design in the communication domain relies on practical channel conditions, and the signals are considered man-made representations for reliable propagation. Thus, the optimal input and output representations for DL communication systems remain unknown. The inputs are represented as binary or one-hot vectors in previous studies but are not limited to the two schemes without verification of optimality. The selection of loss functions and training strategies, as well as the training of the DL system whether on a fixed SNR or a range, present other topics worthy of investigation in future research.
	
	\subsubsection{From simulation to implementation}
	Most of the DL-based algorithms designed for the physical layer of wireless communication systems are still in their simulation stages. To the best of our knowledge, only \cite{dorner2017deep} attempts to investigate the implementation of these algorithms. Therefore, researchers still have to improve these algorithms considerably before they can be implemented. First, an authentic set of data from real communication systems or prototype platforms in actual physical environments must be made available to all researchers to help them train their DL architectures on common measured data and compare the performance of different algorithms objectively. Second, the communication channels in the simulation are often generated by certain models. Therefore, DL systems achieve comparable performance because of their impressive expressive capacity. However, the diverse physical channel scenarios are considerably more complex in reality and change over time. Given that the current DL systems are mainly trained offline, their generalization capability must be guaranteed. Designing specialized systems for specific scenarios or general systems that dynamically adapt to VC conditions is also imperative. DL tools for hardware, such as field programmable gate array, must be developed to deploy the DL methods on hardware and achieve fast realization.
	\section{Conclusion}
	\label{conclusion}
	This paper reviews the literature on the application of DL methods to the physical layer of wireless communication systems to replace parts of the conventional communication system or create a new DL-based architecture (i.e., an autoencoder system). Given their excellent expressive capacity and convenient optimization, the DL-based algorithms show competitive performance with less complexity or latency and have potential application in future communication systems where conventional theories are challenged. The application of DL to the physical layer of wireless communication systems presents a promising research area that is far from maturity. Further studies, including solid theoretical analyses, must be conducted and new DL-based architectures must be proposed to implement DL-based ideas in actual communication scenarios.		
	\bibliographystyle{IEEEtran}
	\bibliography{IEEEabrv,myBib}

\begin{thebibliography}{10}
\providecommand{\url}[1]{#1}
\csname url@samestyle\endcsname
\providecommand{\newblock}{\relax}
\providecommand{\bibinfo}[2]{#2}
\providecommand{\BIBentrySTDinterwordspacing}{\spaceskip=0pt\relax}
\providecommand{\BIBentryALTinterwordstretchfactor}{4}
\providecommand{\BIBentryALTinterwordspacing}{\spaceskip=\fontdimen2\font plus
\BIBentryALTinterwordstretchfactor\fontdimen3\font minus
  \fontdimen4\font\relax}
\providecommand{\BIBforeignlanguage}[2]{{%
\expandafter\ifx\csname l@#1\endcsname\relax
\typeout{** WARNING: IEEEtran.bst: No hyphenation pattern has been}%
\typeout{** loaded for the language `#1'. Using the pattern for}%
\typeout{** the default language instead.}%
\else
\language=\csname l@#1\endcsname
\fi
#2}}
\providecommand{\BIBdecl}{\relax}
\BIBdecl

\bibitem{O2017An}
\BIBentryALTinterwordspacing
T.~J. O'Shea and J.~Hoydis. (2017) An introduction to deep learning for the
  physical layer. [Online]. Available: \url{https://arxiv.org/abs/1702.00832,
  preprint.}
\BIBentrySTDinterwordspacing

\bibitem{larsson2014massive}
E.~G. Larsson, O.~Edfors, F.~Tufvesson, and T.~L. Marzetta, ``Massive {MIMO}
  for next generation wireless systems,'' \emph{{IEEE} Commun. Mag.}, vol.~52,
  no.~2, pp. 186--195, Feb. 2014.

\bibitem{Prelcic17ArXiv}
\BIBentryALTinterwordspacing
{N. Gonz\'{a}lez-Prelcic, A. Ali, V. Va, and R. W. Heath Jr}. (2017) Millimeter
  wave communication with out-of-band information. [Online]. Available:
  \url{https://arxiv.org/abs/1703.10638, preprint.}
\BIBentrySTDinterwordspacing

\bibitem{farsad2017detection}
\BIBentryALTinterwordspacing
N.~Farsad and A.~Goldsmith. (2017) Detection algorithms for communication
  systems using deep learning. [Online]. Available:
  \url{https://arxiv.org/abs/1705.08044, preprint.}
\BIBentrySTDinterwordspacing

\bibitem{jeon2016supervised}
\BIBentryALTinterwordspacing
Y.-S. Jeon, S.-N. Hong, and N.~Lee. (2016) Supervised-learning-aided
  communication framework for massive {MIMO} systems with low-resolution
  {ADC}s. [Online]. Available: \url{https://arxiv.org/abs/1610.07693,
  preprint.}
\BIBentrySTDinterwordspacing

\bibitem{tanweiqiang}
W.~Tan, S.~Jin, C.~K. Wen, and Y.~Jing, ``Spectral efficiency of mixed-{ADC}
  receivers for massive {MIMO} systems,'' \emph{IEEE Access}, vol.~4, pp.
  7841--7846, Sep. 2016.

\bibitem{samuel2017deep}
\BIBentryALTinterwordspacing
N.~Samuel, T.~Diskin, and A.~Wiesel. (2017) Deep {MIMO} detection. [Online].
  Available: \url{https://arxiv.org/abs/1706.01151, preprint.}
\BIBentrySTDinterwordspacing

\bibitem{dorner2017deep}
\BIBentryALTinterwordspacing
{S. D{\"o}rner, S. Cammerer, J. Hoydis, and S. {ten} Brink}. (2017) Deep
  learning-based communication over the air. [Online]. Available:
  \url{https://arxiv.org/abs/1707.03384, preprint.}
\BIBentrySTDinterwordspacing

\bibitem{Goodfellow-et-al-2016}
I.~Goodfellow, Y.~Bengio, and A.~Courville, \emph{Deep Learning}.\hskip 1em
  plus 0.5em minus 0.4em\relax MIT Press, 2016,
  \url{http://www.deeplearningbook.org}.

\bibitem{challita2017proactive}
\BIBentryALTinterwordspacing
U.~Challita, L.~Dong, and W.~Saad. (2017) Proactive resource management in
  {LTE-U} systems: A deep learning perspective. [Online]. Available:
  \url{https://arxiv.org/abs/1702.07031, preprint.}
\BIBentrySTDinterwordspacing

\bibitem{daniels2010adaptation}
R.~C. Daniels, C.~M. Caramanis, and R.~W. Heath, ``Adaptation in
  convolutionally coded {MIMO-OFDM} wireless systems through supervised
  learning and {SNR} ordering,'' \emph{{IEEE} Trans. Veh. Technol.}, vol.~59,
  no.~1, pp. 114--126, Jan. 2010.

\bibitem{pulliyakode2017reinforcement}
\BIBentryALTinterwordspacing
S.~K. Pulliyakode and S.~Kalyani. (2017) Reinforcement learning techniques for
  outer loop link adaptation in {4G/5G} systems. [Online]. Available:
  \url{https://arxiv.org/abs/1708.00994, preprint.}
\BIBentrySTDinterwordspacing

\bibitem{vieira2017deep}
\BIBentryALTinterwordspacing
J.~Vieira, E.~Leitinger, M.~Sarajlic, X.~Li, and F.~Tufvesson. (2017) Deep
  convolutional neural networks for massive {MIMO} fingerprint-based
  positioning. [Online]. Available: \url{https://arxiv.org/abs/1708.06235,
  preprint.}
\BIBentrySTDinterwordspacing

\bibitem{fehske2005new}
A.~Fehske, J.~Gaeddert, and J.~H. Reed, ``A new approach to signal
  classification using spectral correlation and neural networks,'' in
  \emph{Proc. IEEE Int. Symp. New Frontiers in Dynamic Spectrum Access Networks
  (DYSPAN)}, 2005, pp. 144--150.

\bibitem{azzouz1996modulation}
E.~E. Azzouz and A.~K. Nandi, ``Modulation recognition using artificial neural
  networks,'' in \emph{Proc. Automatic Modulation Recognition of Communication
  Signals}, 1996, pp. 132--176.

\bibitem{ibukahla1997neural}
M.~Ibukahla, J.~Sombria, F.~Castanie, and N.~J. Bershad, ``Neural networks for
  modeling nonlinear memoryless communication channels,'' \emph{{IEEE} Trans.
  Commun.}, vol.~45, no.~7, pp. 768--771, Jul. 1997.

\bibitem{sjoberg1995nonlinear}
J.~Sj{\"o}berg, Q.~Zhang, L.~Ljung, A.~Benveniste, B.~Delyon, P.-Y. Glorennec,
  H.~Hjalmarsson, and A.~Juditsky, ``Nonlinear black-box modeling in system
  identification: a unified overview,'' \emph{Automatica}, vol.~31, no.~12, pp.
  1691--1724, Oct. 1995.

\bibitem{bruck1989neural}
J.~Bruck and M.~Blaum, ``Neural networks, error-correcting codes, and
  polynomials over the binary n-cube,'' \emph{{IEEE} Trans. Inf. Theory},
  vol.~35, no.~5, pp. 976--987, Sep. 1989.

\bibitem{ortuno1992error}
I.~Ortuno, M.~Ortuno, and J.~Delgado, ``Error correcting neural networks for
  channels with {G}aussian noise,'' in \emph{Proc. IJCNN International Joint
  Conference on Neural Networks}, vol.~4, 1992, pp. 295--300.

\bibitem{wen2015channel}
C.~K. Wen, S.~Jin, K.~K. Wong, J.~C. Chen, and P.~Ting, ``Channel {E}stimation
  for {M}assive {MIMO} {U}sing {G}aussian-{M}ixture {B}ayesian {L}earning,''
  \emph{{IEEE} Trans. Wireless Commun.}, vol.~14, no.~3, pp. 1356--1368, Mar.
  2015.

\bibitem{chen1990adaptive}
S.~Chen, G.~Gibson, C.~Cowan, and P.~Grant, ``Adaptive equalization of finite
  non-linear channels using multilayer perceptrons,'' \emph{Elsevier Signal
  processing}, vol.~20, no.~2, pp. 107--119, Jun. 1990.

\bibitem{cid1996digital}
J.~Cid-Sueiro and A.~R. Figueiras-Vidal, ``Digital equalization using modular
  neural networks: an overview,'' in \emph{Proc. Signal Processing in
  Telecommunications}.\hskip 1em plus 0.5em minus 0.4em\relax Springer, 1996,
  pp. 337--345.

\bibitem{ibnkahla2000applications}
M.~Ibnkahla, ``Applications of neural networks to digital communications--a
  survey,'' \emph{Elsevier Signal processing}, vol.~80, no.~7, pp. 1185--1215,
  Jul. 2000.

\bibitem{jiang2017machine}
C.~Jiang, H.~Zhang, Y.~Ren, Z.~Han, K.-C. Chen, and L.~Hanzo, ``Machine
  learning paradigms for next-generation wireless networks,'' \emph{{IEEE}
  Wireless Commun.}, vol.~24, no.~2, pp. 98--105, Apr. 2017.

\bibitem{hornik1989multilayer}
K.~Hornik, M.~Stinchcombe, and H.~White, ``Multilayer feedforward networks are
  universal approximators,'' \emph{Neural Networks}, vol.~2, no.~5, pp.
  359--366, 1989.

\bibitem{nachmani2016learning}
E.~Nachmani, Y.~Be'ery, and D.~Burshtein, ``Learning to decode linear codes
  using deep learning,'' in \emph{Proc. Communication, Control, and Computing
  (Allerton)}, 2016, pp. 341--346.

\bibitem{nachmani2017rnn}
\BIBentryALTinterwordspacing
E.~Nachmani, E.~Marciano, D.~Burshtein, and Y.~Be'ery. (2017) {RNN} decoding of
  linear block codes. [Online]. Available:
  \url{https://arxiv.org/abs/1702.07560, preprint.}
\BIBentrySTDinterwordspacing

\bibitem{gruber2017deep}
{T. Gruber and S. Cammerer and J. Hoydis and S. ten Brink}, ``On deep
  learning-based channel decoding,'' in \emph{Proc. of CISS}, 2017, pp. 1--6.

\bibitem{cammerer2017scaling}
\BIBentryALTinterwordspacing
T.~Gruber, S.~Cammerer, J.~Hoydis, and S.~ten Brink. (2017) Scaling deep
  learning-based decoding of polar codes via partitioning. [Online]. Available:
  \url{https://arxiv.org/abs/1702.06901, preprint.}
\BIBentrySTDinterwordspacing

\bibitem{nachmani2017deep}
\BIBentryALTinterwordspacing
E.~Nachmani, E.~Marciano, L.~Lugosch, W.~J. Gross, D.~Burshtein, and Y.~Beery.
  (2017) Deep learning methods for improved decoding of linear codes. [Online].
  Available: \url{https://arxiv.org/abs/1706.07043, preprint.}
\BIBentrySTDinterwordspacing

\bibitem{liang2017iterative}
\BIBentryALTinterwordspacing
F.~Liang, C.~Shen, and F.~Wu. (2017) An iterative {BP-CNN} architecture for
  channel decoding. [Online]. Available: \url{https://arxiv.org/abs/1707.05697,
  preprint.}
\BIBentrySTDinterwordspacing

\bibitem{ye2017power}
\BIBentryALTinterwordspacing
H.~Ye, G.~Y. Li, and B.-H.~F. Juang. (2017) Power of deep learning for channel
  estimation and signal detection in {OFDM} systems. [Online]. Available:
  \url{https://arxiv.org/abs/1708.08514, preprint.}
\BIBentrySTDinterwordspacing

\bibitem{neumann2017learning}
\BIBentryALTinterwordspacing
D.~Neumann, T.~Wiese, and W.~Utschick. (2017) Learning the {MMSE} channel
  estimator. [Online]. Available: \url{https://arxiv.org/abs/1707.05674,
  preprint.}
\BIBentrySTDinterwordspacing

\bibitem{o2017deep}
\BIBentryALTinterwordspacing
T.~J. O'Shea, T.~Erpek, and T.~C. Clancy. (2017) Deep learning based {MIMO}
  communications. [Online]. Available: \url{https://arxiv.org/abs/1707.07980,
  preprint.}
\BIBentrySTDinterwordspacing

\bibitem{nandi1998algorithms}
A.~K. Nandi and E.~E. Azzouz, ``Algorithms for automatic modulation recognition
  of communication signals,'' \emph{{IEEE} Trans. Commun.}, vol.~46, no.~4, pp.
  431--436, Apr. 1998.

\bibitem{xin2004automatic}
{X.-Z. Lv, P. Wei, and X.-C. Xiao}, ``Automatic identification of digital
  modulation signals using high order cumulants,'' \emph{Electronic Warfare},
  vol.~6, p. 001, Jun. 2004.

\bibitem{azzouz1996recognition}
A.~K. Nandi and E.~E. Azzouz, ``Automatic analogue modulation recognition,''
  \emph{Elsevier Signal processing}, vol.~46, no.~2, pp. 211--222, Oct. 1995.

\bibitem{caid1990neural}
W.~R. Caid and R.~W. Means, ``Neural network error correcting decoders for
  block and convolutional codes,'' in \emph{Proc. {IEEE} Global Telecommun.
  Conf. (GLOBECOM)}, 1990, pp. 1028--1031.

\bibitem{di1991use}
A.~Di~Stefano, O.~Mirabella, G.~Di~Cataldo, and G.~Palumbo, ``On the use of
  neural networks for {H}amming coding,'' in \emph{Proc. IEEE Int. Sympoisum on
  Circuits and Systems}, 1991, pp. 1601--1604.

\bibitem{wang1996artificial}
X.-A. Wang and S.~B. Wicker, ``An atificial neural net {V}iterbi decoder,''
  \emph{{IEEE} Trans. Commun.}, vol.~44, no.~2, pp. 165--171, Feb. 1996.

\bibitem{dimnik2009improved}
I.~Dimnik and Y.~Be'ery, ``Improved random redundant iterative {HDPC}
  decoding,'' \emph{{IEEE} Trans. Commun.}, vol.~57, no.~7, Jul. 2009.

\bibitem{Mousavi2015CS}
{A. Mousavi, A. B. Patel, and R. G. Baraniuk}, ``A deep learning approach to
  structured signal recovery,'' in \emph{Proc. Communication, Control, and
  Computing (Allerton)}, 2015, pp. 1336--1343.

\bibitem{Mousavi2017CS}
{A. Mousavi and R. G. Baraniuk}, ``Learning to invert: Signal recovery via deep
  convolutional networks,'' in \emph{Proc. Int. Conf. Acoustics, Speech and
  Signal Processing}, 2017, pp. 2272--2276.

\bibitem{Lohit2017CS}
\BIBentryALTinterwordspacing
S.~Lohit, K.~Kulkarni, R.~Kerviche, P.~Turaga, and A.~Ashok. (2017)
  Convolutional neural networks for non-iterative reconstruction of
  compressively sensed images. [Online]. Available:
  \url{https://arxiv.org/abs/1708.04669, preprint.}
\BIBentrySTDinterwordspacing

\end{thebibliography}

\begin{IEEEbiography}[{\includegraphics[width=1in,height=1.25in,clip,keepaspectratio]{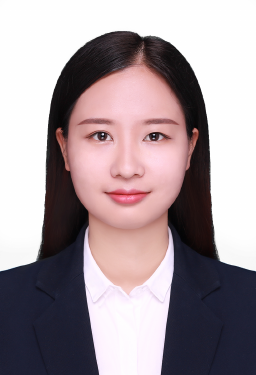}}]{Tianqi Wang}
Tianqi Wang was born in Jiangsu, China, in 1993. She received the B.S. degree from Nanjing University of Science and Technology, Nanjing, China in 2016. She is currently working toward the M.S. degree with the School of Information Science and Engineering, Southeast University. Her main research interests include deep learning application in communication and massive MIMO systems.
\end{IEEEbiography}
\vspace	{-45pt}
\begin{IEEEbiography}[{\includegraphics[width=1in,height=1.25in,clip,keepaspectratio]{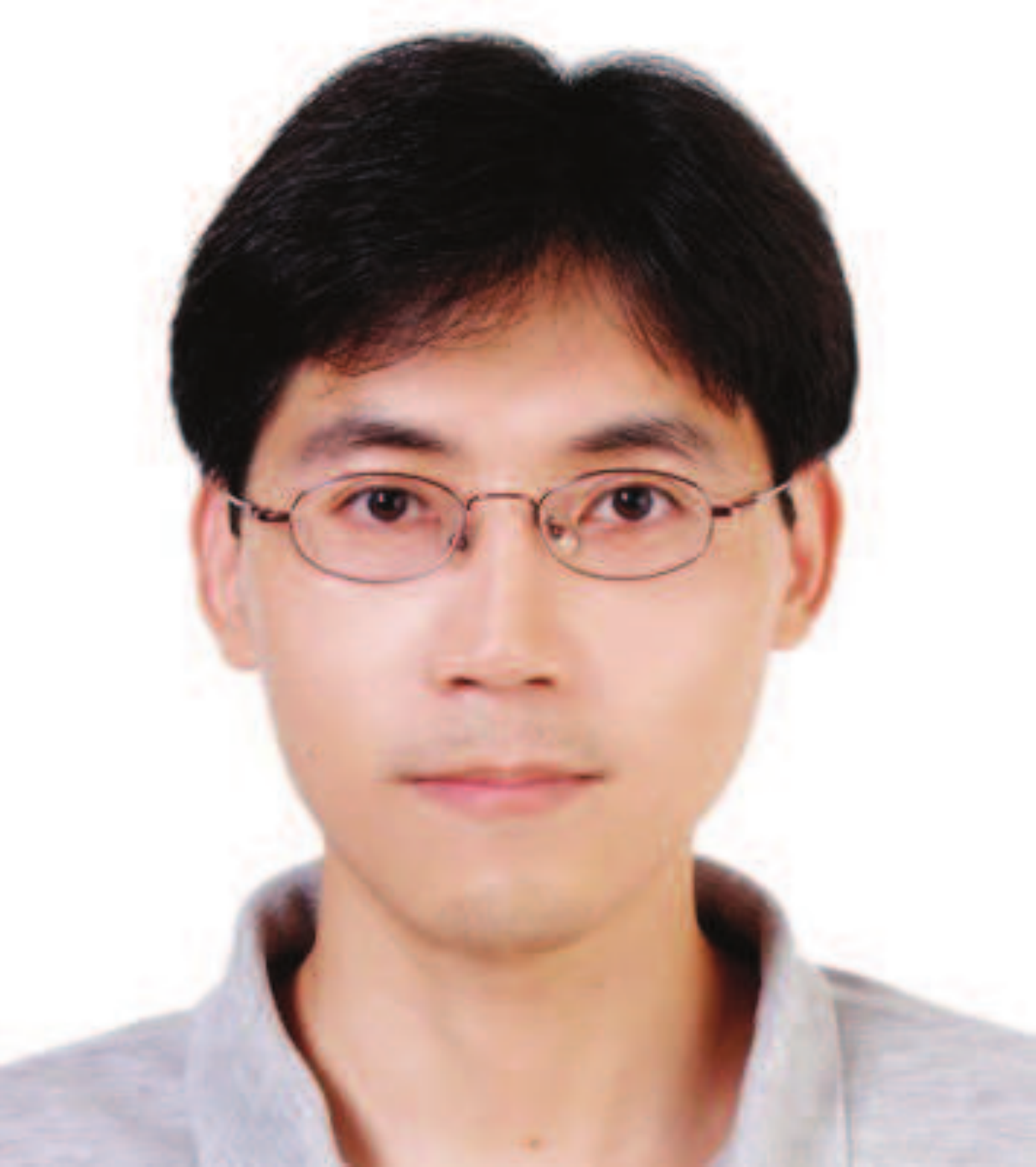}}]{Chao-Kai Wen}
Chao-Kai Wen (S'00-M'04) received the Ph.D. degree from the Institute of Communications Engineering, National Tsing Hua University, Taiwan, in 2004. He was with Industrial Technology Research Institute, Hsinchu, Taiwan and MediaTek Inc., Hsinchu, Taiwan, from 2004 to 2009. He is currently an Associate Professor of the Institute of Communications Engineering, National Sun Yat-sen University, Kaohsiung, Taiwan. His research interests center around the optimization in wireless multimedia networks.
\end{IEEEbiography}
\vspace	{-45pt}
\begin{IEEEbiography}[{\includegraphics[width=1in,height=1.25in,clip,keepaspectratio]{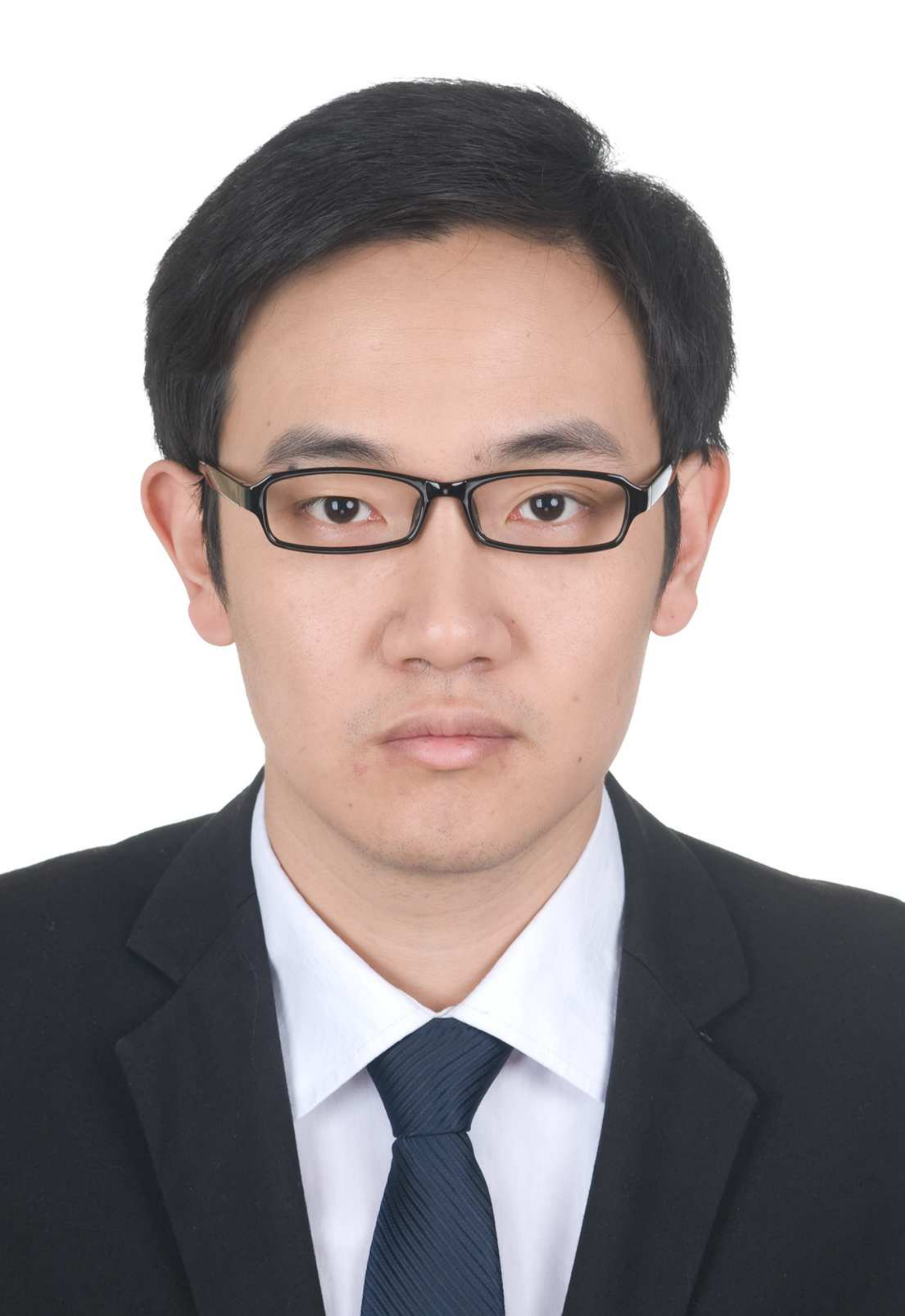}}]{Hangqing Wang}
Hanqing Wang (S’16) received the B.S. degree in communications engineering from Anhui University, Hefei, China, in 2013, and the M.S. degree in information and communications engineering from Southeast University, Nanjing, China, in 2016, where he is currently pursuing the Ph.D. degree in information and communications engineering with the School of Information Science and Engineering. His current research interests include detection and estimation theory, compressive sensing, and their application to communication systems with hardware imperfection and nonlinear distortion.
\end{IEEEbiography}	
\vspace	{-45pt}
\begin{IEEEbiography}[{\includegraphics[width=1in,height=1.25in,clip,keepaspectratio]{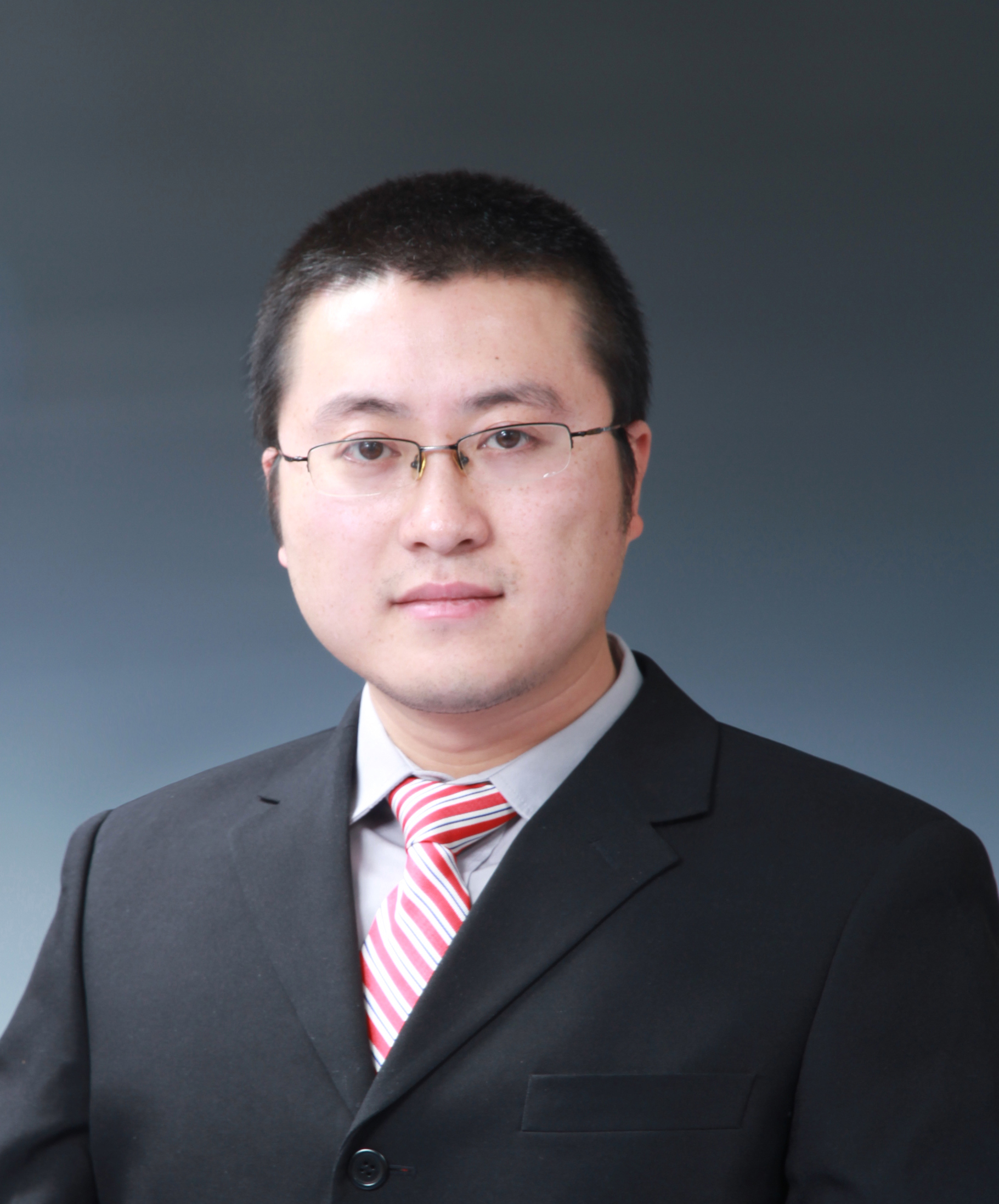}}]{Feifei Gao}
Feifei Gao received the B.Eng. degree from Xi'an Jiaotong University, Xi'an, China in 2002, the M.Sc. degree from McMaster University, Hamilton, ON, Canada in 2004, and the Ph.D. degree from National University of Singapore, Singapore in 2007. He was a Research Fellow with the Institute for Infocomm Research (I2R), A*STAR, Singapore in 2008 and an Assistant Professor with the School of Engineering and Science, Jacobs University, Bremen, Germany from 2009 to 2010. In 2011, he joined the Department of Automation, Tsinghua University, Beijing, China, where he is currently an Associate Professor.
Prof. Gao's research areas include communication theory, signal processing for communications, array signal processing, and convex optimizations, with particular interests in MIMO techniques, multi-carrier communications, cooperative communication, and cognitive radio networks. He has authored/ coauthored more than 100 refereed IEEE journal papers and more than 100 IEEE conference proceeding papers, which have been cited more than 4000 times from Google Scholar.
Prof. Gao has served as an Editor of IEEE Transactions on Wireless Communications, IEEE Communications Letters, IEEE Signal Processing Letters, IEEE Wireless Communications Letters, International Journal on Antennas and Propagations, and China Communications. He has also served as the symposium co-chair for 2015 IEEE Conference on Communications (ICC), 2014 IEEE Global Communications Conference (GLOBECOM), 2014 IEEE Vehicular Technology Conference Fall (VTC), 2018 IEEE Vehicular Technology Conference Spring,as well as Technical Committee Members for many other IEEE conferences.
\end{IEEEbiography}	
\vspace	{-45pt}
\begin{IEEEbiography}[{\includegraphics[width=1in,height=1.25in,clip,keepaspectratio]{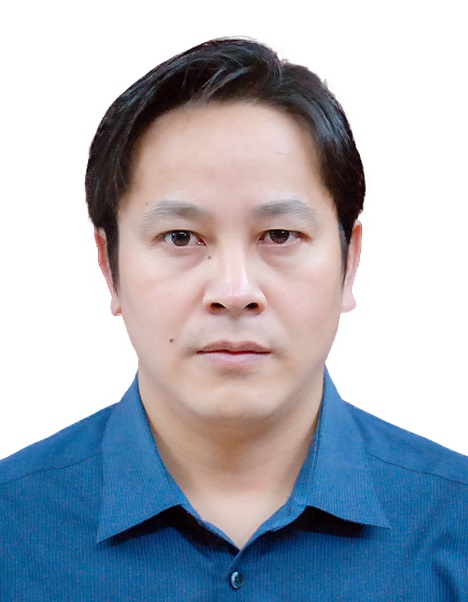}}]{Tao Jiang}
Tao Jiang is currently a Distinguished Professor in the School of Electronics Information and Communications, Huazhong University of Science and Technology, Wuhan, P. R. China. He received Ph.D. degree in information and communication engineering from Huazhong University of Science and Technology, Wuhan, P. R. China, in April 2004. From Aug. 2004 to Dec. 2007, he worked in some universities, such as Brunel University and University of Michigan-Dearborn, respectively. He has authored or co-authored about 300 technical papers in major journals and conferences and 9 books/chapters in the areas of communications and networks. He served or is serving as symposium technical program committee membership of some major IEEE conferences, including INFOCOM, GLOBECOM, and ICC, etc.. He was invited to serve as TPC Symposium Chair for the IEEE GLOBECOM 2013, IEEEE WCNC 2013 and ICCC 2013. He is served or serving as associate editor of some technical journals in communications, including in IEEE Transactions on Signal Processing, IEEE Communications Surveys and Tutorials, IEEE Transactions on Vehicular Technology, IEEE Internet of Things Journal, and he is the associate editor-in-chief of China Communications, etc.. He is a recipient of the NSFC for Distinguished Young Scholars Award in 2013. He was awarded as the Most Cited Chinese Researchers announced by Elsevier in 2014, 2015 and 2016.
\end{IEEEbiography}	
\vspace	{-45pt}
\begin{IEEEbiography}[{\includegraphics[width=1in,height=1.25in,clip,keepaspectratio]{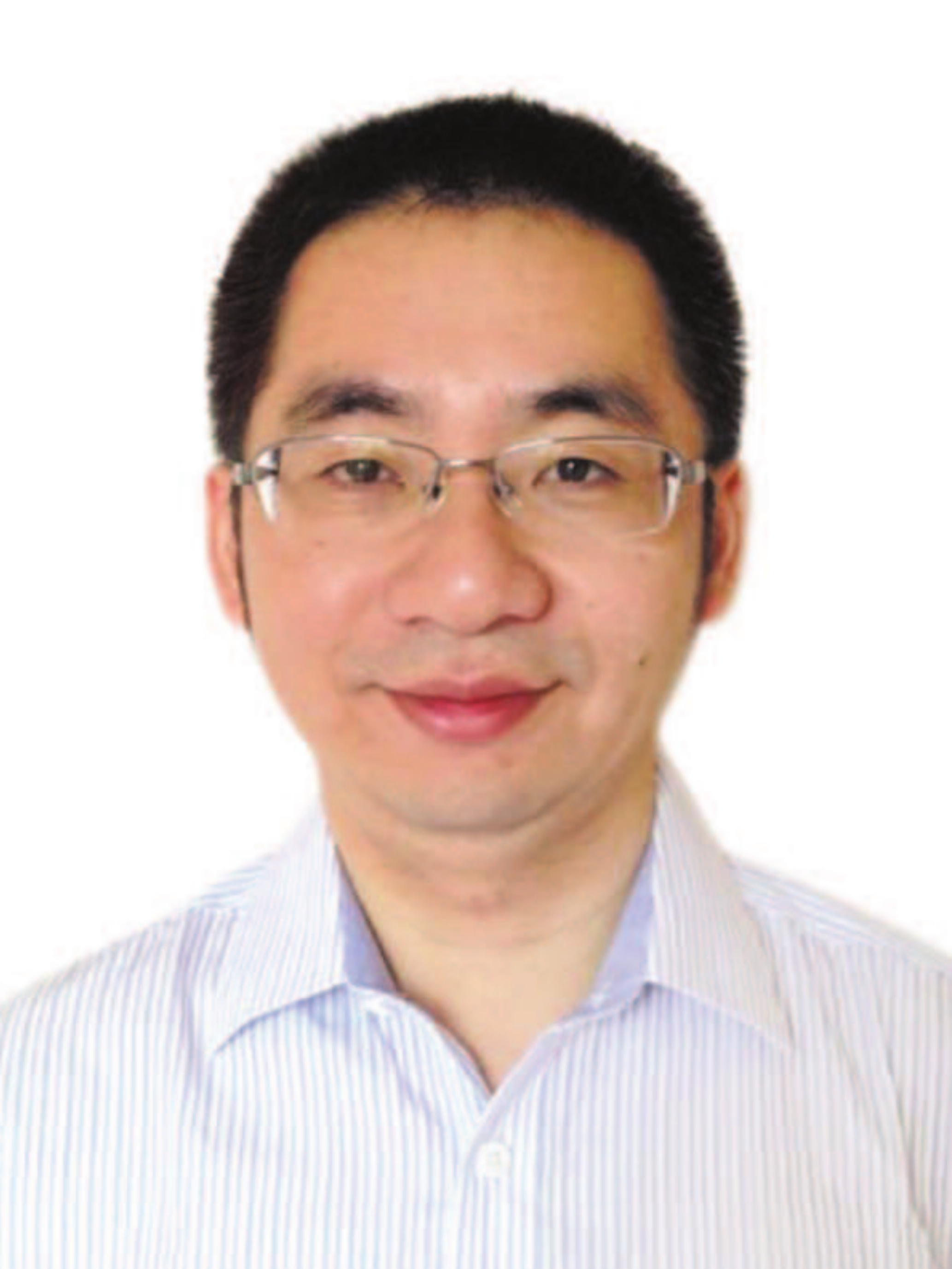}}]{Shi Jin}
Shi Jin (S'06--M'07) received the B.S. degree in communications engineering from Guilin University of Electronic Technology, Guilin, China, in 1996; the M.S. degree from Nanjing University of Posts and Telecommunications, Nanjing, China, in 2003; and the Ph.D. degree in communications and information systems from Southeast University, Nanjing, in 2007. From June 2007 to October 2009, he was a Research Fellow with the Adastral Park Research Campus, University College London, London, U.K. He is currently with the faculty of the National Mobile Communications Research Laboratory, Southeast University. His research interests include space-time wireless communications, random matrix theory, and information theory. Dr. Jin serves as an Associate Editor for the IEEE Transactions on Wireless Communications, the IEEE Communications Letters, and IET Communications. He and his coauthors received the 2010 Young Author Best Paper Award by the IEEE Signal Processing Society and the 2011 IEEE Communications Society Stephen O. Rice Prize Paper Award in the field of communication theory.
\end{IEEEbiography}
\end{document}